\documentclass{aa}  

\usepackage{graphicx}
\usepackage{txfonts}
\usepackage{hyperref}
\hypersetup{
    colorlinks=true,
    linkcolor=blue,
    citecolor=blue,
    urlcolor=blue,
    pdftitle={A Universal Method for Solar Filament Detection from H-alpha Observations using Semi-supervised Deep Learning},
    pdfauthor={Diercke et al.}
}

\usepackage{amsmath}
\usepackage{amssymb}
\usepackage{multirow}
\usepackage{natbib}
\usepackage[rightcaption]{sidecap}
\usepackage{units}
\usepackage{url}
\usepackage[normalem]{ulem}
\usepackage{xcolor}

\newcommand\arcdeg{\mbox{$^\circ$}}
\newcommand\arcsecdot{\mbox{$^{\prime\prime}$}\hspace{-0.15cm}.\,} 

\graphicspath{{./}{figures/}}

\begin{document}

   \title{A Universal Method for Solar Filament Detection from H$\alpha$ Observations using Semi-supervised Deep Learning}
   \titlerunning{A Universal Method for Solar Filament Detection}
   \authorrunning{Diercke et al.}

   \author{Andrea Diercke 
          \inst{1,2,3,4}
          \and
          Robert Jarolim\inst{5, 6} 
          \and
          Christoph Kuckein\inst{7,8,3, 9} 
          \and
          Sergio J. Gonz\'alez Manrique\inst{1,7,8,10} 
          \and
          Marco Ziener\inst{11}
          \and
          Astrid M. Veronig\inst{5,12} 
          \and
          Carsten Denker\inst{3} 
          \and
          Werner Pötzi\inst{5,12} %
          \and
          Tatiana Podladchikova\inst{13} 
          \and
          Alexei A. Pevtsov\inst{2} 
          }

   \institute{
            Institut für Sonnenphysik (KIS), 
            Sch\"oneckstr. 6,
            79104 Freiburg, Germany \\
            \email{diercke@leibniz-kis.de}
        \and 
            National Solar Observatory (NSO), 
             3665 Discovery Drive,
             Boulder, CO, USA, 80303
        \and 
            Leibniz-Institut f\"ur Astrophysik Potsdam (AIP), 
             An der Sternwarte 16, 
             14482 Potsdam, Germany
        \and 
            Universit\"at Potsdam,
            Institut f\"ur Physik und Astronomie,
            Karl-Liebknecht-Stra\ss{}e 24/25,
            14476 Potsdam, Germany
        \and 
            University of Graz,
            Institute of Physics,
            Universit\"atsplatz 5,
            8010 Graz, Austria 
        \and 
            High Altitude Observatory (HAO), 
            3090 Center Green Drive, 
            Boulder, CO, USA, 80301 \\
            \email{rjarolim@ucar.edu}
        \and 
            Instituto de Astrof\'{i}sica de Canarias (IAC), 
            V\'{i}a L\'{a}ctea s/n, 38205 La Laguna, Tenerife, Spain
        \and 
            Departamento de Astrof\'{\i}sica, Universidad de La Laguna
            38205, La Laguna, Tenerife, Spain
        \and 
            Max-Planck-Institut f\"ur Sonnensystemforschung, 
            Justus-von-Liebig-Weg 3, 
            37077 G\"ottingen, Germany
        \and 
            Astronomical Institute, 
            Slovak Academy of Sciences, 
            05960 Tatransk\'a Lomnica, Slovak Republic
        \and 
            Applied AI Company Limited (AAICO),
            C29 - C11-Building 2 Al Khaleej St,
            Al Muntazah - Zone 1, 
            Abu Dhabi, UAE
        \and 
            University of Graz, Kanzelh\"ohe Observatory for Solar and Environmental Research, Kanzelh\"ohe 19, 9521 Treffen am Ossiacher See, Austria
        \and 
            Skolkovo Institute of Science and Technology, 
            Bolshoy Boulevard 30, bld. 1, 
            Moscow 121205, Russia
             }

   \date{Received October 18, 2023; accepted XXXX XX, XXXX}

 
  \abstract
   {Filaments are omnipresent features in the solar atmosphere. Their location, properties and time evolution can provide important information about changes in solar activity and assist the operational space weather forecast. Therefore, filaments have to be identified in full disk images and their properties extracted from these images. Manual extraction is tedious and takes too much time; extraction with morphological image processing tools produces a large number of false-positive detections. Automatic object detection, segmentation, and extraction in a reliable manner allows us to process more data in a shorter time. The Chromospheric Telescope (ChroTel), Tenerife, Spain, the Global Oscillation Network Group (GONG), and the Kanzelh\"{o}he Observatory for Solar and Environmental Research (KSO), Austria, provide  regular full-disk  observations of the Sun in the core of the chromospheric H$\alpha$ absorption line. We present a deep learning method that provides reliable extractions of solar filaments from H$\alpha$ filtergrams.  First, we train the object detection algorithm YOLOv5 with labeled filament data of ChroTel H$\alpha$ filtergrams. We use the trained model to obtain bounding-boxes from the full GONG archive. In a second step, we apply a semi-supervised training approach, where we use the bounding boxes of filaments, to learn a pixel-wise classification of solar filaments with u-net. Here, we make use of the increased data set size which avoids overfitting of spurious artifacts from the generated training masks. Filaments are predicted with an accuracy of 92\%. With the resulting filament segmentations, physical parameters such as the area or tilt angle can be easily determined and studied. This we demonstrate in one example, where we determine the rush-to-the pole for Solar Cycle~24 from the segmented GONG images. In a last step, we apply the filament detection to H$\alpha$ observations from KSO which demonstrates the general applicability of our method to H$\alpha$ filtergrams.}


   \maketitle
%

\section{Introduction}

Solar filaments are structures of dense, cool plasma, which reach from the chromosphere into the corona, stabilized by the magnetic field. They appear above and along the polarity inversion line (PIL), which separates large areas of opposite magnetic polarities \citep{Martin1998a, Mackay2010}. Filaments can be very dynamic objects, which change their appearance in only a few hours. If the magnetic field is destabilized, the filaments can erupt, e.g., as coronal mass ejections (CMEs), and the plasma stored in them is ejected into space \citep{Wang2020, Kuckein2020}. We differentiate between three  kinds of filaments \citep{Bruzek.Durrant1977}: i) active region filaments, which are smaller, more dynamic, and rooted at large magnetic field concentrations of active regions; ii) quiescent filaments, which appear at all latitudes and are usually larger and more stable; and iii) the category of intermediate filaments, where filaments are subsumed, if they do not fit to the aforementioned classes \citep{Mackay2010}. A special type of quiescent filaments are polar crown filaments, which appear at high-latitudes above 50\arcdeg\ \citep{Leroy1983}. Systematic filament studies have been performed across all filament classes mainly utilizing H$\alpha$ full-disk filtergrams \citep[e.g., statistical studies by][]{Hao2015, Poetzi2015, Diercke2019b, Chatzistergos2023}. The key ingredient of all these studies is an effective extraction of the location, length, width, and other statistical properties of filaments from full-disk observations.

Automatic filament detection was attempted in several studies mostly employing intensity threshold supplemented by other information  \citep[e.g., presence/absence of magnetic field neutral line, see ][]{Shih.Kowalski2003,Qu2005,Scholl.etal2008,Joshi.etal2010,Karachik.Pevtsov2014}. \citet{Xu2018} resorted to a visual inspection of the location of polar crown filaments, because they could not find a satisfactory method to extract all filaments automatically. \citet{Hao2015} applied an adaptive threshold, which is based on the \citet{Otsu1979} threshold, but with a modification, which takes into account the mean intensity on the solar disk to improve the automatic detection of filaments. \citet{Diercke2019b} extracted filaments automatically at all latitudes from synoptic charts based on H$\alpha$ full-disk images by applying morphological image processing techniques. To avoid the extraction of sunspots from the data, they excluded small-scale features, which eliminated small-scale filaments as well. Moreover, the algorithm encountered problems extracting  filaments at the poles, in particular, because in the geometrical conversion to an equidistant grid with heliographic coordinates the filaments were stretched resulting in some artifacts in the maps.

Extracting filaments from H$\alpha$ full-disk observations using conventional image processing tools is challenging due to the complexity of distinguishing them from other dark features, such as sunspots. \citet{Zhu2019, Zhu2020} presented an approach to identify filaments from full-disk H$\alpha$ images \citep{Denker1999} obtained at the  Big Bear Solar Telescope (BBSO), using the fully convolutional neural network (FCN) u-net \citep{Ronneberger2015}, which was modified to enable image segmentation to extract filaments. The authors train the network with semi-manual computed ground-truth using image manipulation software and classical image processing. The results show an efficient segmentation compared to classical image processing, with reduced false detection of small-scale dark objects. Nonetheless, the  exclusion of sunspots is not effective in this method. \citet{Liu2021} used the same data set but with additional data from the Huairou Solar Observing Station of the National Astronomical Observatory, China \citep{Suo2020}. They extended with their results the comprehensive data set from \citet{Zhu2019}. Another recent approach to segment filaments is presented by  \citet{Ahmadzadeh2019}, who used an off-the-shelf model, Mask Region-Based Convolutional Neural Network \citep[R-CNN, ][]{Girshick2014, He2017} and applied it to BBSO H$\alpha$ full-disk data. The training data are created from maps, which are also used as input for the  Heliophysics Events Knowledgebase \citep[HEK, ][]{Hurlburt2012}. In the data, non-filament regions are removed with classical morphological image processing, whereby the accuracy is estimated to be about 72\% compared to manually labeled data \citep{Ahmadzadeh2019}. The segmentation results are comparable to the ground-truth labels and the network identified filaments that were missing in the original HEK database. In the approach of \citet{Guo2022}, the BBSO data from 2010 to 2015 is labeled with an online tool. They include also data, which is not corrected for limb darkening and randomly vary the brightness in the training data set to improve the robustness of the segmentation method. The authors employed the Conditional Convolutions for Instance Segmentation model \citep[CondInst][]{Tian2020} for segmentation, which adeptly manages input data of various shapes and demonstrates a reliable performance across images of varying quality.

\bigskip

\begin{figure}
\includegraphics[width=1\columnwidth]{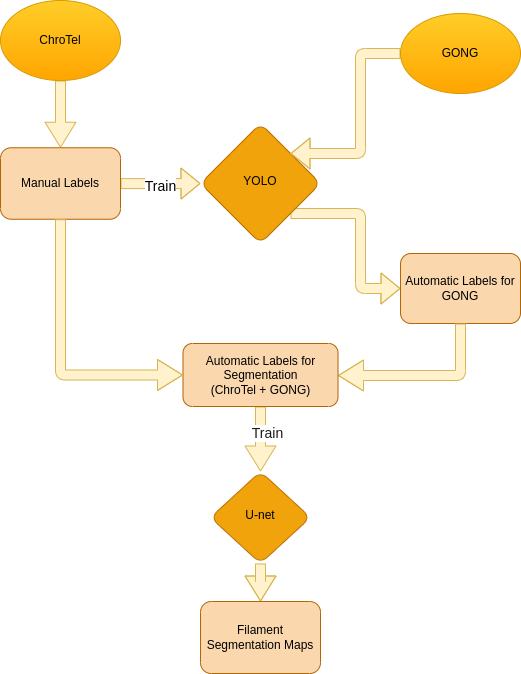}
\caption{Schematic outline of filament detection and segmentation with neural networks.}
\label{fig:ml_schema}
\end{figure}

Extraction of filaments for statistical studies can be tedious. Morphological image processing is very effective, but a differentiation with other dark features on the Sun is challenging. In this study, we use a well-established deep neural network, i.e., YOLOv5 \citep{Jocher2020Yolov5}, with the labeled data of the Chromospheric Telescope \citep[ChroTel, ][]{Bethge2011} to detect filaments in H$\alpha$ full-disk filtergrams (Sect.~\ref{sec:yolo}). For the study of solar filaments (e.g., angle, length), a pixel-wise classification is needed. We apply a threshold per filament bounding-box to provide a segmentation mask. However, in the case of extended bounding boxes, this approach may introduce artifacts and can lead to the detection of sunspots. We mitigate this shortcoming by training an additional segmentation model \citep[i.e., u-net][]{Ronneberger2015}. Here, we first apply the trained YOLOv5 model to the H$\alpha$ filtergrams of the  Global Oscillation Network Group \citep[GONG, ][]{Harvey1996} with a 4-h cadence (17\,413 observations), from which we generate our segmentation maps for the segmentation model training. We assume that invalid segmentations appear as noise in the training data and the detection of filaments is more robustly learned by the u-net model (c.f., Sect. \ref{sec:unet}). In other words, we use a semi-supervised training approach to achieve filament segmentations with reduced false detections (e.g, sunspots). The schematic outline of the present study is displayed in Figure~\ref{fig:ml_schema}. We compare our results to the classical morphological image processing from \citet{Diercke2019b}, which demonstrates a clear improvement (Sect.~\ref{sec:morph}). Afterwards the results from the different steps are evaluated (Sect.~\ref{sec:eval}). We further apply our trained model to H$\alpha$ filtergrams from the  Kanzelh\"ohe Observatory for Solar and Environmental Research \citep[KSO, ][]{Otruba2003, Poetzi2015, Jarolim2020, Poetzi2021} to assess our method for H$\alpha$ observations from different instruments (Sect.~\ref{sec:kso}). This demonstrates that our method can extract filaments from various full-disk H$\alpha$ instruments and data source. In the end, we obtain the masks predicted by the segmentation model, where we can extract filaments and use them as input for a statistical study, similar to the study by \citet{Diercke2019b}. These filaments could then be used, e.g., for statistical studies with multiple applications (c.f., Sect.~\ref{sec:rushpole}) or simply as input data for global databases such as HEK.


%
\section{Data} \label{sec:data}
%

\begin{figure}
\includegraphics[width=1.\columnwidth]{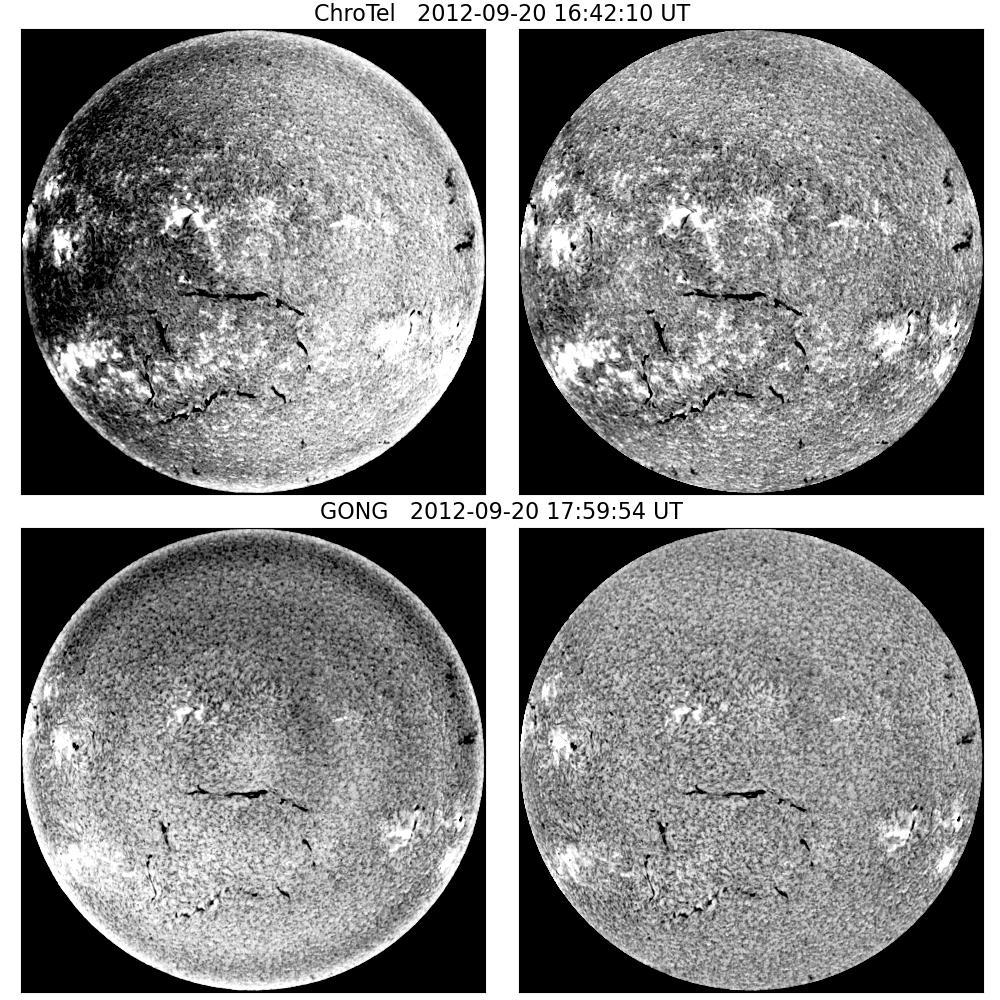}
\caption{Intensity correction of H$\alpha$ ChroTel (top) and GONG (bottom) filtergrams: center-to-limb-variation corrected image (left) and intensity corrected image (right).}
\label{fig:intcorr}
\end{figure}

Our study employs H$\alpha$ data from three sources: ChroTel, GONG, and Kanzelh\"ohe Observatory. In order to train a neural network to perform object detection of filaments, we label these filaments in H$\alpha$ full-disk filtergrams. For this purpose, we use the data set of ChroTel \citep{Kentischer2008, Bethge2011}. The 10-cm aperture telescope mounted at the terrace of the VTT  \citep{vonderLuehe1998} observes in three different wavelengths H$\alpha$, \mbox{Ca\,\textsc{ii}}~K, and \mbox{He\,\textsc{i}}~$\lambda$10830\,\AA\ with a cadence of 3\,min. The observations are carried out with three Lyot filters, whereby the H$\alpha$ filter has a FWHM of 0.5\,\AA. The regular ChroTel observations started in 2012 and it continued to acquire data until 2020. The ChroTel data set contains 1056~days of observations until September 2020, whereby on each day the qualitatively best filtergram is selected using the Median Filter-Gradient Similarity \citep[MFGS, ][]{Deng2015, Denker2018} method. In this method, the similarity between the original image and the median-filtered image is evaluated. The basic data reduction includes dark- and flat-field correction, as well as rotation to the solar north and geometric correction of solar images with an oval appearance, which is caused by the low elevation of the Sun in the early mornings or late evenings. Furthermore, the images are normalized to the median intensity of the solar disk, are limb darkening corrected, and the off-limb region is truncated. The Lyot filter introduced a non-uniform intensity variation (Figure~\ref{fig:intcorr}, top row), which is corrected by approximation with Zernike polynomials \citep{Shen2018}. Further details of the image processing are given in \citet{Shen2018} and \citet{Diercke2019b}. All filtergrams are scaled to a solar radius of $r = 1000$\,pixels, which results in an image scale of about  0\arcsecdot96~pixel$^{-1}$ with an image size of $2000\times2000$\,pixels.

\begin{figure}
\includegraphics[width=1.\columnwidth]{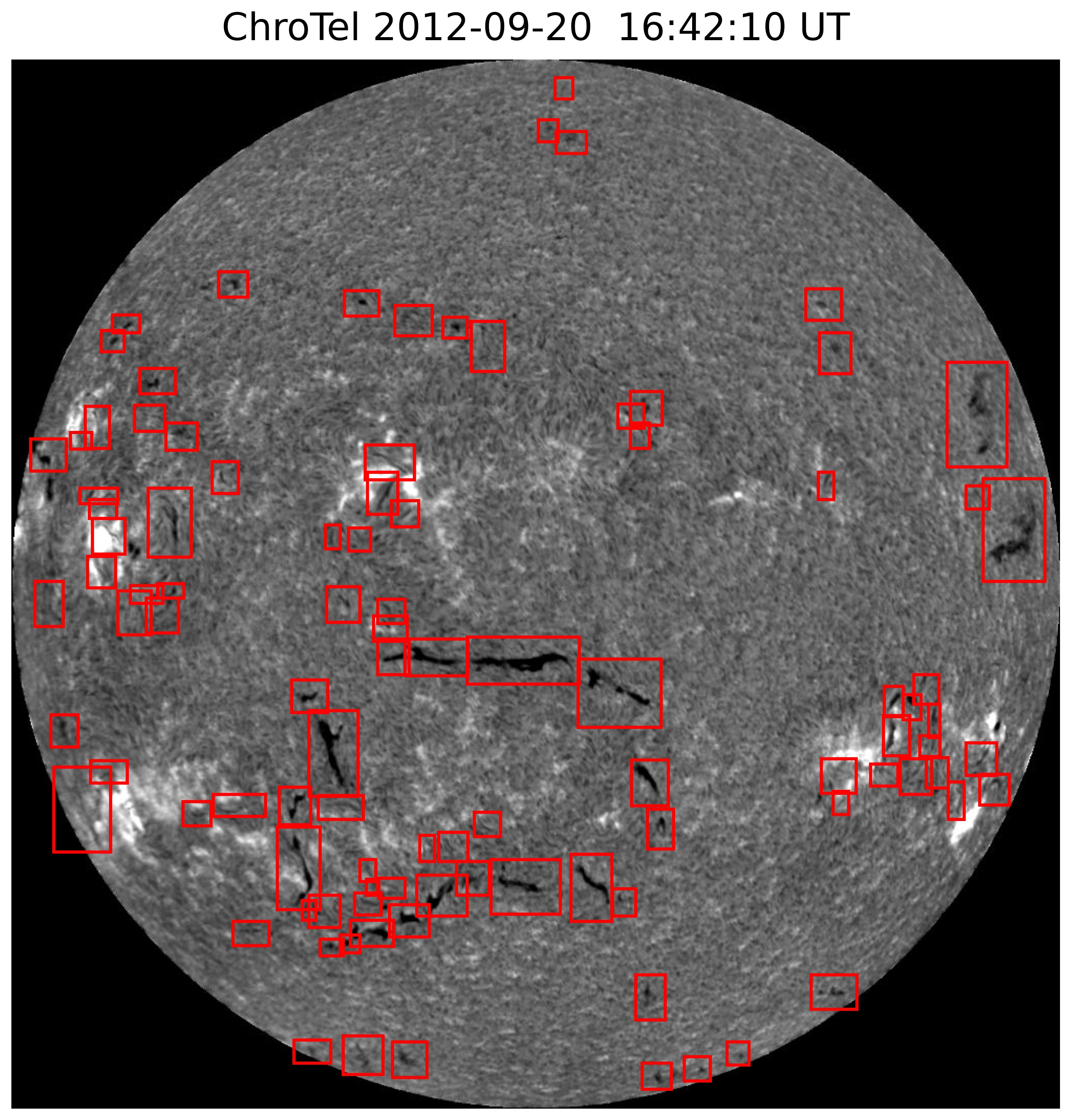}
\caption{H$\alpha$ ChroTel filtergram with the manually labeled bounding boxes (red) for 2012-09-20. These bounding boxes define the ground truth and input data for the object detection algorithm.}
\label{fig:chrotel_label}
\end{figure}

Three observers manually labeled one image of each day from the entire ChroTel data set between 2012 and 2018, which includes 955 observing days. For each image, we  labeled each filament by determining the upper left and bottom right corner to construct a rectangular bounding box containing the filament. In some cases the filament is split in several parts, whereby we labeled each part individually with a bounding box. The labeled data cover observations from the maximum and minimum of Solar Cycle~24. The labels of the bounding boxes and re-scaled images with a resolution of 1024$\times$1024\,pixels are used as ground-truth input data for the training of the deep neural network. An example of such an input image with the corresponding labels is displayed in Figure~\ref{fig:chrotel_label}. For each red bounding box we save the information on the central coordinate of the box, its width and height, and the information of the type of object (class). In this study, we only have one class, i.e.,  filaments,  which has the class identifier 0.

\bigskip

The Global Oscillation Network Group \citep[GONG, ][]{Harvey1996, Hill2018} is operated by the National Solar Observatory (NSO) Integrated Synoptic Program (NISP) since 1995 with the goal to acquire nearly continuous observations of oscillations on the solar surface. The network is comprised of six identical stations situated at different longitudes around the world, which ensures average duty cycle of about 93\% \citep[][]{Jain2021}. GONG instruments were designed to take full disk observations of Doppler shifts and the line-of-sight magnetograms in the Ni {\rm I} 6768\,\AA spectral line. In 2010, the GONG stations were upgraded to take also observations in the core of H$\alpha$ spectral line (last system was deployed in December 2010). The H$\alpha$ instrument uses existing GONG light feed. The GONG objective lens is 80\,mm diameter and 1000\,mm focal length. It is vignetted by the entrance turret to about 70\,mm effective aperture, which defines the theoretical resolution of the GONG system. A polarizing beamsplitter sends light in the H$\alpha$ wavelength through the Daystar Quantum PE~0.4\,\AA\ mica etalon filter. The filter is placed in a telecentric beam using two re-imaging lenses ($f=450$\,mm and $f=800$\,mm) before the filter. The filters use a unique dual-heater system developed by Daystar Filters LLC to compensate for a persistent index gradient that shifts center wavelength across surface of mica etalons. The filter aperture (32\,mm) is matched to the GONG entrance pupil. Following the filter, two lenses (positive achromat, $f=300$\,mm, and  negative achromat, $f=-100$\,mm) form a Galilean telescope that creates a full disk solar image on a Digital Video Camera Co. DVC-4000AM 2K$\times$2K interline uncooled CCD camera \citep{Harvey2011}. Image resolution is limited by diffraction, atmospheric seeing and high-order wavefront errors  in the filter. The diffraction-limited spatial resolution of the final image is a little over 2$^{\prime\prime}$. It is sampled by the CCD camera at about 1$^{\prime\prime}$ per pixel\footnote{GONG Doppler velocity and magnetograph leg is further limited in its theoretical resolution to an equivalent aperture of 2.8\,cm or about 5$^{\prime\prime}$ sampled at 2\arcsecdot5 per pixel}.

The exposure time is adjusted automatically at each site to keep quiet disk center at 20\% of (14-bit) dynamic range. First, a pre-exposure of four images are taken with a default, running average, exposure time. The images are added together, and dark corrected. Flat, smear, and sky-brightness correction are also applied. We note that the sky-brightness correction is an average brightness of the square regions of the image corners. This average brightness is then subtracted and a corresponding scale factor is multiplied to bring the image back to a 20\% dynamic range; the purpose for this correction is to compensate for known Daystar filter outgassing between maintenance visits, as well as correcting for local atmospheric conditions. The final resulting pre-exposure image is used to calculate the exposure time for the science image. Ten seconds later, the next 4 images are taken with the new pre-exposure time,  applying  the same dark, flat, smear, and sky-brightness corrections, and the process repeats.

The cadence for each site is 60 seconds  (relative to the fixed GPS time due to Universal Time leap second adjustments), but because of an (intentionally) scheduled 20 second offset in time acquisition between the neighboring sites, an overall network cadence of up to 20\,s can be achieved \citep{Jain2021}. The images are  2048$\times$2048\,pixels in size. They are processed at each GONG site, compressed via the JPEG2000 (J2K) algorithm (with a slight loss to meet GONG sites’ bandwidth limitations), transferred to NISP Data Center in Boulder, Colorado, uncompressed, and become available to public within one minute of acquisition. Because of the light feed design, the H$\alpha$ images rotate during the day relative to the fixed filter and camera orientations. As part of the processing, the images are digitally de-rotated using solar ephemeris and the information about the position of GONG turrets for each site during the day. This correction is only approximate.

The GONG data is used as additional training set for the segmentation neural network. Therefore, we download GONG data with a cadence of 4\,h between 2010~June~01 and 2021~February~23, which results in a total number of 17\,413 images. We use the Image-quality Assessment method described in \citet{Jarolim2020} to filter observations, which suffer from atmospheric degradation and exclude them from our data set. This provides a more reliable filtering of degraded images than could be achieved through quality metrics based on pixel distributions (e.g., contrast). The final GONG data set contains then a total number of 16\,759 images. For all our data we use the same pre-processing to mitigate instrumental differences and to enhance the image contrast. This includes the reduction of the image resolution to 1024$\times$1024\,pixels, correction of the center-to-limb-variation, and correction of additional intensity variations using Zernike polynomials (Figure~\ref{fig:intcorr}, bottom row), as described for ChroTel data. Finally, we clip values to [0.8, 1.3], followed by normalizing the data to the interval [$-1$, 1].

\bigskip

We use additional H$\alpha$ observations from the Kanzelh\"ohe Observatory \citep[KSO, ][]{Otruba2003, Poetzi2015, Poetzi2021} to validate the object detection and segmentation algorithm with an additional  data source. KSO is observing in three wavelengths: white-light, \mbox{Ca\,\textsc{ii}}~K, and H$\alpha$. The H$\alpha$ observations are part of the Global H$\alpha$ Network \citep[GHN, ][]{Steinegger2000}, providing  H$\alpha$ full-disk images on a daily basis with a cadence of about 1\,min with an image scale of about 1\arcsec~pixel$^{-1}$.

%
\section{Methods} \label{sec:methods}
%

\subsection{Object Detection} \label{sec:objdet}

In this section, we give a brief introduction on object detection with deep neural networks \citep{Elgendy2020} and specifically the object detection algorithm You Only Look Once \cite[YOLO, ][]{Redmon2016a, Redmon2016b, Redmon2018, Jocher2020Yolov5}. Object detection algorithms localize an object in an image, classify each object based on the trained examples, and try to determine an optimal bounding box around the object. 

One common evaluation method for object detection algorithms is the Intersection over Union (IoU), which can be defined for each bounding box. The fraction of the overlap and the union of the bounding box of the ground-truth $B_{\mathrm{GT}}$ and the bounding box of the prediction $B_{\mathrm{Pred}}$ is calculated as \citep{Elgendy2020}:
\begin{equation}
 \mathrm{IoU} = \frac{B_{\mathrm{GT}} \cap B_{\mathrm{Pred}}}{B_{\mathrm{GT}} \cup B_{\mathrm{Pred}}}. \label{eq:iou}
\end{equation}
The IoU defines the correct number of predictions, the True Positives (TP). Therefore, a threshold is used when a bounding box is defined as a correct prediction. The threshold for a true positive detection of a bounding box is above an IoU of 0.5, otherwise the bounding box is considered as a false positive (FP) prediction. If an object is not detected, we have a false negative prediction (FN). Along these, we define the precision, recall, and accuracy \citep{Elgendy2020}:
\begin{align}
  \mathrm{Recall} =  \frac{\mathrm{TP}}{\mathrm{TP}+\mathrm{FN}}  \label{eq:prec}    \\
  \mathrm{Precision} =  \frac{\mathrm{TP}}{\mathrm{TP}+\mathrm{FP}}. \label{eq:recall} \\
  \mathrm{Accuracy} =  \frac{\sum{\mathrm{GT}}-\mathrm{FN}}{\sum{\mathrm{GT}}}  \label{eq:acc},
\end{align}
whereby GT refers to the number of objects contained in the ground truth. By plotting the precision and recall values, we can define the Precision-Recall curve and the area below this curve is the Average Precision (AP). The mean over all classes in the problem is finally the mAP, which is always given at a certain threshold for the IoU score. An mAP@0.5 describes the threshold for an IoU score of 0.5.

The object detection algorithm YOLO is a single-stage detector, which means that the object class and the objectness score are predicted in the same step. This algorithm is capable of performing detections in real-time, where the filament prediction is typically in the range of milliseconds on a modern GPU, which is far below the selected 60\,s cadence of GONG. In YOLO the region proposal step is omitted and instead the image is split into grids, where for each cell a bounding box is predicted. Non-maximum suppression allows us to extract from the large number of candidates the final bounding box. There are three detection layers for detection on large-, medium-, and small-scale. For this purpose, the image is divided into three grids of different sizes. We use YOLOv5\footnote{\href{https://github.com/ultralytics/yolov5}{https://github.com/ultralytics/yolov5}} \citep{Jocher2020Yolov5} by Ultralytics\footnote{\href{https://www.ultralytics.com/}{https://www.ultralytics.com/}}, which uses the Python library for deep learning applications \texttt{PyTorch} \citep{Paszke2019}. The model architecture is built-up of three main parts: the backbone, the neck, and the head of the model. The backbone mainly extracts the important features from the input image. The neck is responsible to detect features in the images in different scales and to generalize the model, which is performed in YOLOv5 with a feature pyramid. The model head performs the final detection. It decides on the bounding boxes, the class probabilities, and objectness of the detection. 

For the prediction of filament bounding boxes we use the YOLOv5 model. To use the model for solar physics applications, we have to use a customized training of YOLOv5 from scratch, whereby the weights of the algorithm are randomly initialized, hence no transfer learning is used. For the model architecture we use the large configuration \citep[c.f, ][]{Jocher2020Yolov5}. For the training we use  the full-disk ChroTel  H$\alpha$ filtergrams with a resolution of $1024 \times 1024$ pixels. The data set is split in a test set, a training set, and a validation set. The training set is the largest with 681 images (70\%) and the validation set contains 84\,images (10\%). The test set contains the remaining 192 images (20\%), whereby we select for each year one continuous block of 20\% of images for the test set. The image and the corresponding file containing the coordinates of each bounding box are the input for the training of the neural network.

\begin{figure*}
\includegraphics[width=1\textwidth]{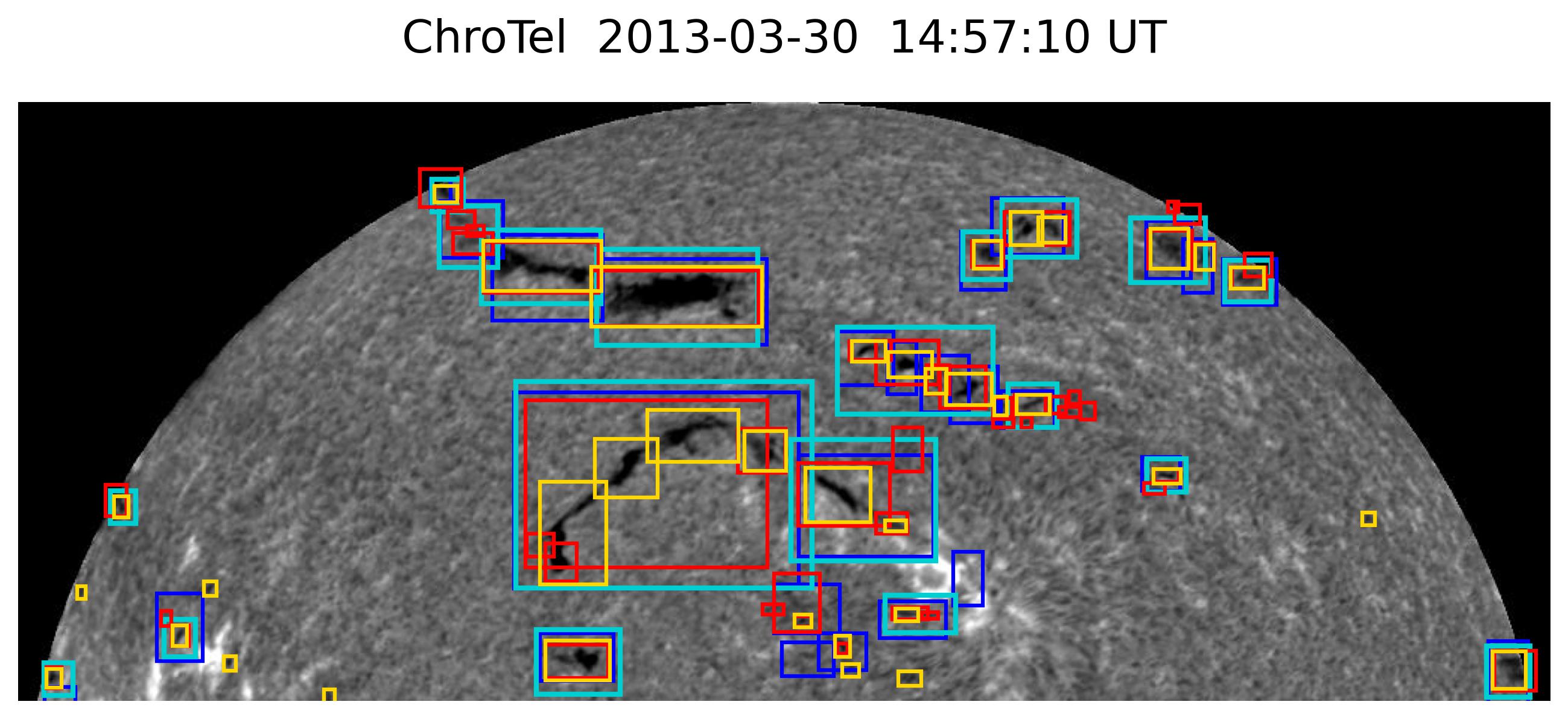}
\caption{Comparison of the bounding boxes from the ground truth (blue boxes), YOLO predictions (light blue boxes), the bounding boxes created from u-net segmentation (red boxes) and bounding boxes created from segmentations using morphological image processing (yellow boxes).}
\label{fig:comp_morph_bb_all}
\end{figure*}

\subsection{Segmentation} \label{sec:seg}

While bounding boxes allow for a convenient labeling, the precise study of solar filaments requires a pixel-wise identification. As a baseline approach we use a global intensity threshold of 1$\sigma$ and select only pixels enclosed in bounding boxes. The problem with this approach is that extended filaments and bounding boxes that overlap with sunspots lead to invalid identifications. 

For our image segmentation application we build on a u-net architecture \citep{Ronneberger2015}, which is a frequently employed neural network architecture for semantic segmentation. U-net is a fully convolutional network with 23 convolutional layers. The architecture is u-shaped and is composed of a contracting path and an expansive path. The contracting path contains a series of convolutional layers, each followed by a rectified linear unit (ReLU), which is used to downsample the input maps. The downsampling is accompanied by the doubling of the number of feature channels. In the expansive path, the process is reversed and with each step of the alternating convolutional layer and the ReLU, the maps are upsampled by a $2\times2$ convolution and the number of feature channels is halved. This is a concatenation with the feature maps from the contracting path.

In our study, we aim for a general method that can provide reliable solar filament detection for arbitrary H$\alpha$ observations. We employ a semi-supervised approach where we take advantage of the more robust bounding boxes and the full information of a pixel-wise segmentation. We start by training a YOLOv5 model to predict bounding-boxes, based on our labeled ChroTel observation. The resulting model is used to label the full GONG archive of solar observations at a 4-hour cadence, where we use an equal splitting between the available observing sites.

\begin{figure}
\includegraphics[width=1\columnwidth]{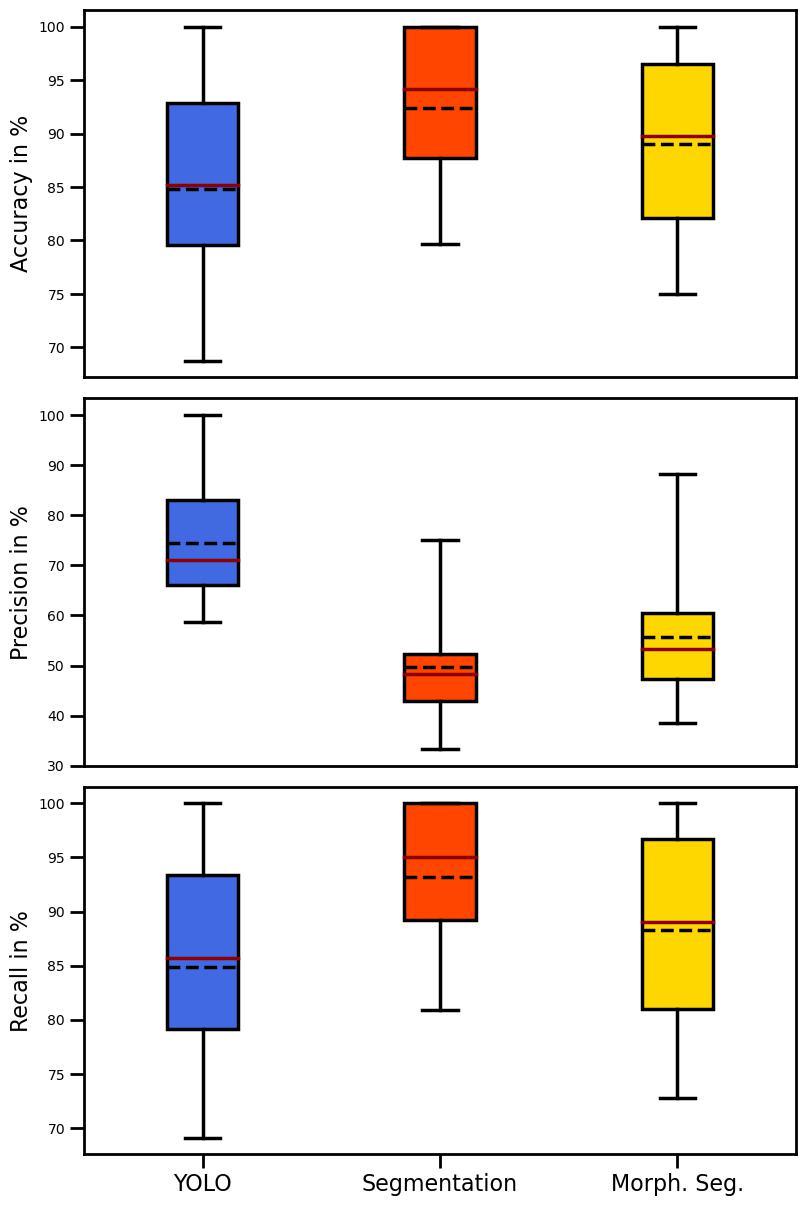}
\caption{Box plots for the accuracy, precision, and recall in percent from YOLO object detection (left, blue box), the segmentation with u-net (middle, red box), and segmentation maps obtained from morphological image processing (right, yellow box). The box is displayed between the 25th and 75th percentile, and whiskers between the 5th and 95th percentile.  The median is marked as a dark red line and the mean as a dashed black line.}
\label{fig:box_morph_all}
\end{figure}

From this we apply our thresholding approach to create a large data set of solar filament segmentations, which we use for training a final u-net model. We note that the baseline thresholding approach inevitably contains invalid pixel-wise classifications (e.g., sunspots). While noisy labels can limit the model performance, we build on the concept that for our extended data set (18\,472~images) the spurious classifications are not fitted by the network. From this we expect that the neural network does not learn to replicate the noise and focuses on task of filament classification, leading to a more reliable segmentation than the simple thresholding approach. We note that training the u-net model only with the manual labels is possible, but does not provide a sufficiently large data set to mitigate misclassifications (e.g., the model tends to classify sunspots as filaments). For the u-net test set, we use the same continuous blocks of GONG and ChroTel observations as described in Sect.~\ref{sec:objdet} for the training of YOLO on ChroTel data. The validation set is the same as for the YOLO training.

\subsection{Setting a baseline - segmentation with morphological image processing} \label{sec:morph}

To set a baseline for the performance of the machine learning algorithm, we compare the segmentation results with results from classical morphological image processing. For this purpose, we use the ChroTel H$\alpha$ filtergrams, which are corrected for intensity variations with Zernike polynomials. These filtergrams are normalized on the median intensity. We use a threshold, which is the median intensity plus additional ten percent of the median intensity \citep[see descriptions in ][]{Diercke2019b, Diercke2022} to create a mask of dark structures, such as filaments for the filtergrams. In addition, we use morphological opening and closing to create coherent contours of the filaments. As last step, we filter out small-scale structures with less than five pixels, to remove non-filament structures.

%
\section{Results}
%

\subsection{Quantitative Evaluation of the Results} \label{sec:eval}

\begin{figure}
\includegraphics[width=1\columnwidth]{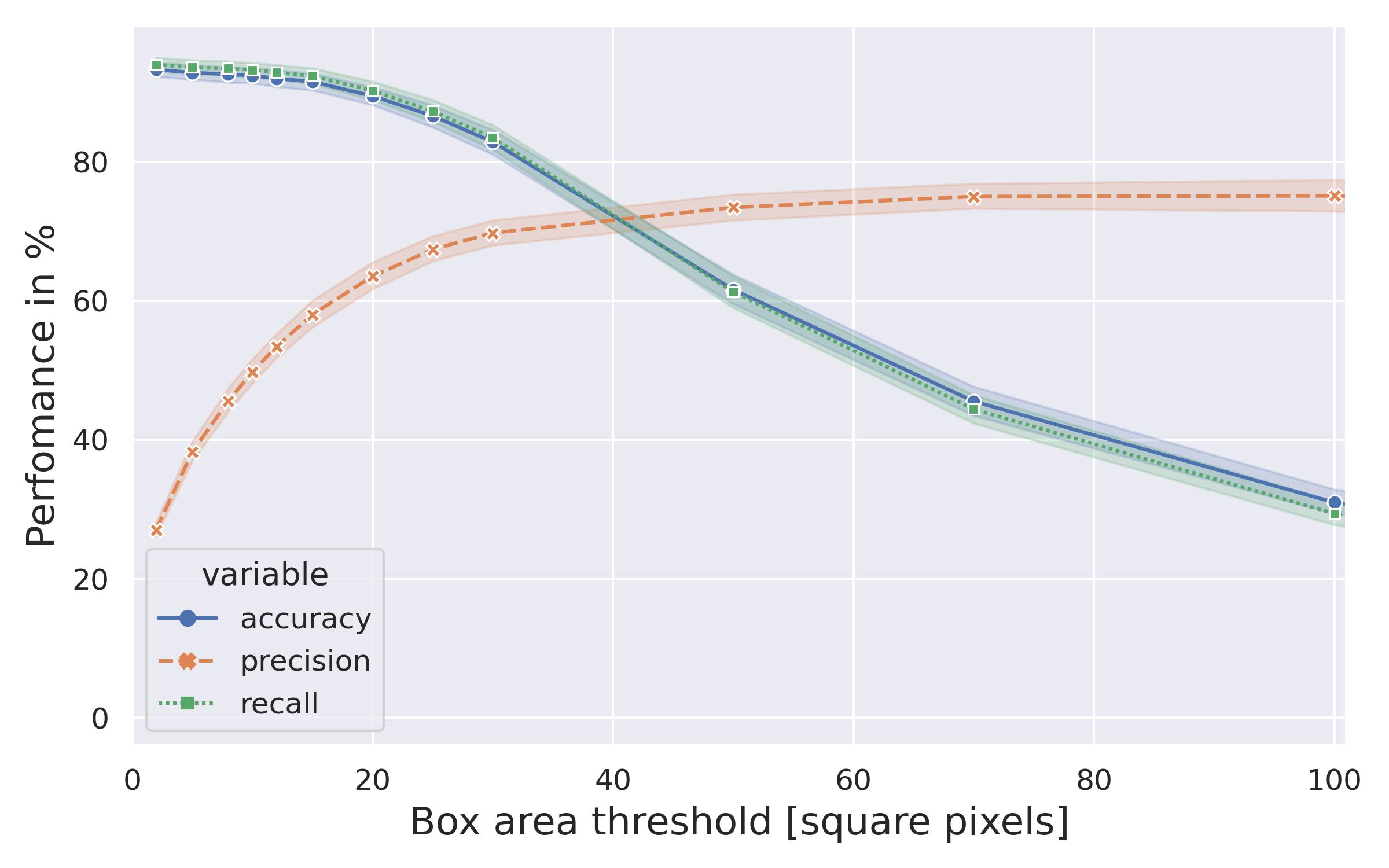}
\caption{Recall, accuracy, and precision in percent for different thresholds for the box size.}
\label{fig:rsize_thres}
\end{figure}

The quantitative evaluation of results involves a comparative analysis between the bounding boxes derived from ground truth annotations and those predicted by YOLO for ChroTel\footnote{Since the selected GONG data is not simultaneously observed with the ChroTel data used to label the ground truth, the location of the filaments is slightly shifted and a quantitative evaluation with the ground truth is not possible.}, alongside bounding boxes generated from segmentation maps produced by u-net and morphological image processing techniques. The data set used for the evaluation is based on all filaments predicted in each image from the test set for ChroTel data. We employ bounding box comparison instead of pixel-wise assessment due to the initial generation of ground truth annotations in the bounding box format, rendering it a more convenient method of evaluation. Furthermore, our study primarily emphasizes the comprehensiveness of detection rather than the precise segmentation of individual pixels. One drawback of this approach lies in the complex handling of overlapping bounding boxes. In Figure~\ref{fig:comp_morph_bb_all}, we display an example with all utilized bounding boxes. The ground truth (light blue boxes) are bounding boxes, which were labeled manually. Since the YOLO results contain the location of the predicted bounding boxes (blue boxes), the comparison is straightforward. For the segmentation results with u-net (red boxes) and morphological image processing (yellow boxes), we have to estimate bounding boxes around the contours, for which we use the \texttt{regionprops} routine from the \texttt{Python scikit-image} library. Since the bounding boxes of the ground truth always contain a certain border around the filament, the YOLO predictions mimic this behaviour. For the artificially created bounding boxes around the u-net and morphological contours, we add a 20\% additional height and width, that the bounding boxes can be better compared with the ground truth. For larger filaments, where the width or height exceeds 50\,pixels, the enlargement is 10\% of the original height and width. During the labelling process, the bounding boxes around larger filaments are tightly encapsulating the filament, in contrast to bounding boxes around smaller filaments. Therefore, we selected two distinct padding values, which were reasonably chosen to fit the bounding boxes from the ground truth. The bounding boxes are used to calculate the IoU as in Eq.~(\ref{eq:iou}), precision as in Eq.~(\ref{eq:prec}) and recall as in Eq.~(\ref{eq:recall}). Thereby, a true positive is considered, when the IoU is above 10\%. With the relatively low IoU score and padded bounding-boxes we aim to mitigate the false classification of differently separated filaments. For example, Figure~\ref{fig:comp_morph_bb_all} shows that elongated filaments are separated into different bounding boxes by the individual methods, where we still consider the separated detections as correct. We note that the correct separation of elongated filaments would require additional information about the magnetic field topology. The results for the accuracy, precision, and recall is displayed as boxplots in Figure~\ref{fig:box_morph_all} and the corresponding statistical values are displayed in Table~\ref{tab:stat}. We refrain from assessing the IoU due to the study's primary focus on prioritizing the reliability of detection over the precise alignment of bounding boxes.

\begin{table*}[]
\caption{Statistical evaluation of the accuracy, precision, and recall for object detection with YOLO, segmentation with u-net, and morphological image processing compared to the ground truth. The values are given in percent.}
\label{tab:stat}
\centering
\begin{tabular}{|l|ccc|ccc|ccc|ccc|}
\hline\hline 
   &  \multicolumn{3}{c|}{YOLO}  & \multicolumn{3}{c|}{u-net}  & \multicolumn{3}{c|}{Morph. Image Proc.}  \\
{} &   Accuracy &  Precision &  Recall &   Accuracy &  Precision &  Recall &   Accuracy &  Precision &  Recall \\
\hline
mean  &          84.8 &           \textbf{74.5} &        84.9 &         \textbf{92.4} &          49.8 &       \textbf{93.2} &           89.0 &            55.7 &         88.3 \\
std   &          10.8 &           12.3 &        11.2 &         7.9 &          12.8 &       7.3 &           8.9 &            14.1 &         9.6 \\
\hline
\end{tabular}
\end{table*}

\begin{figure*}
\includegraphics[width=0.49\textwidth]{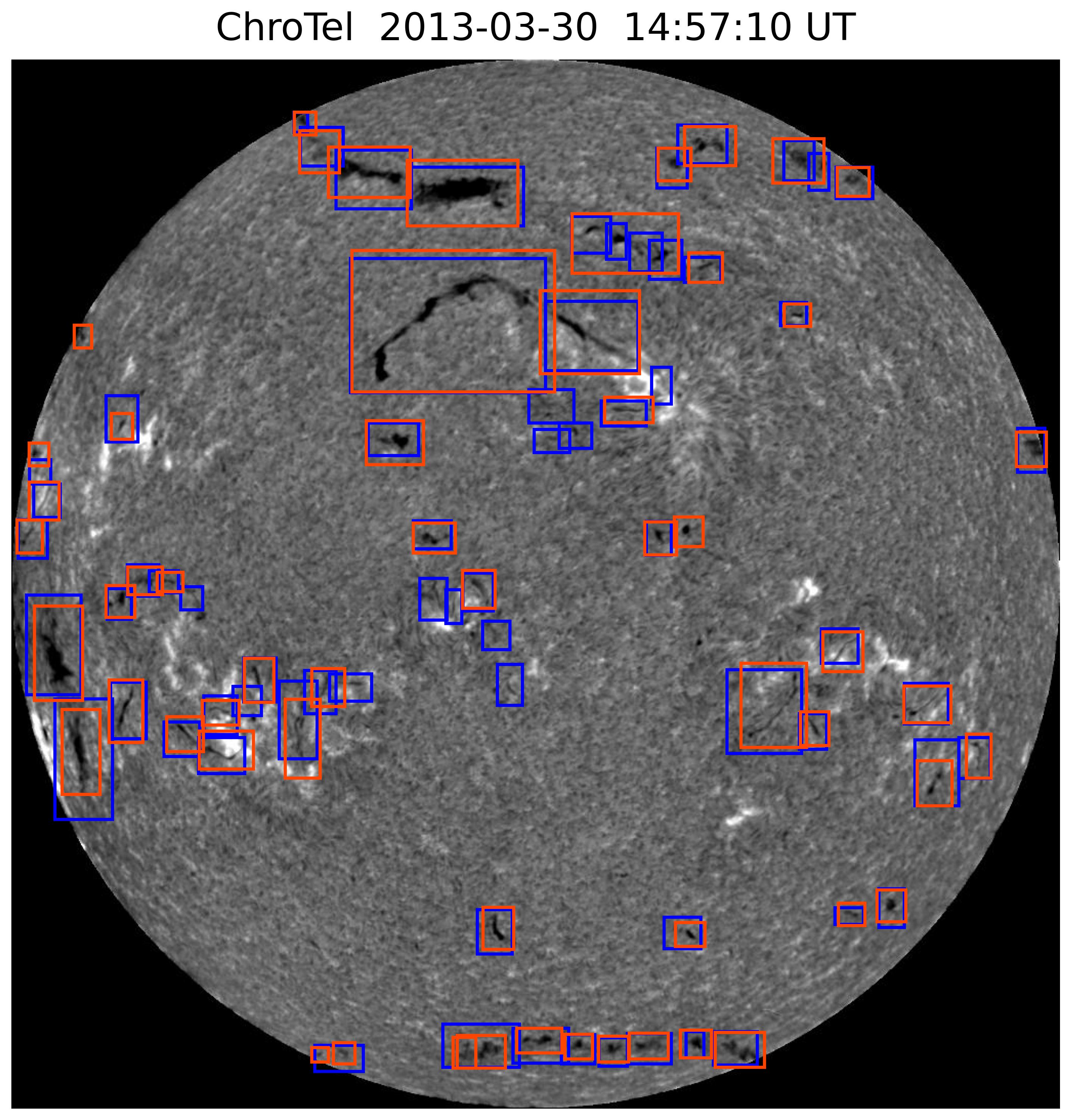}
\includegraphics[width=0.49\textwidth]{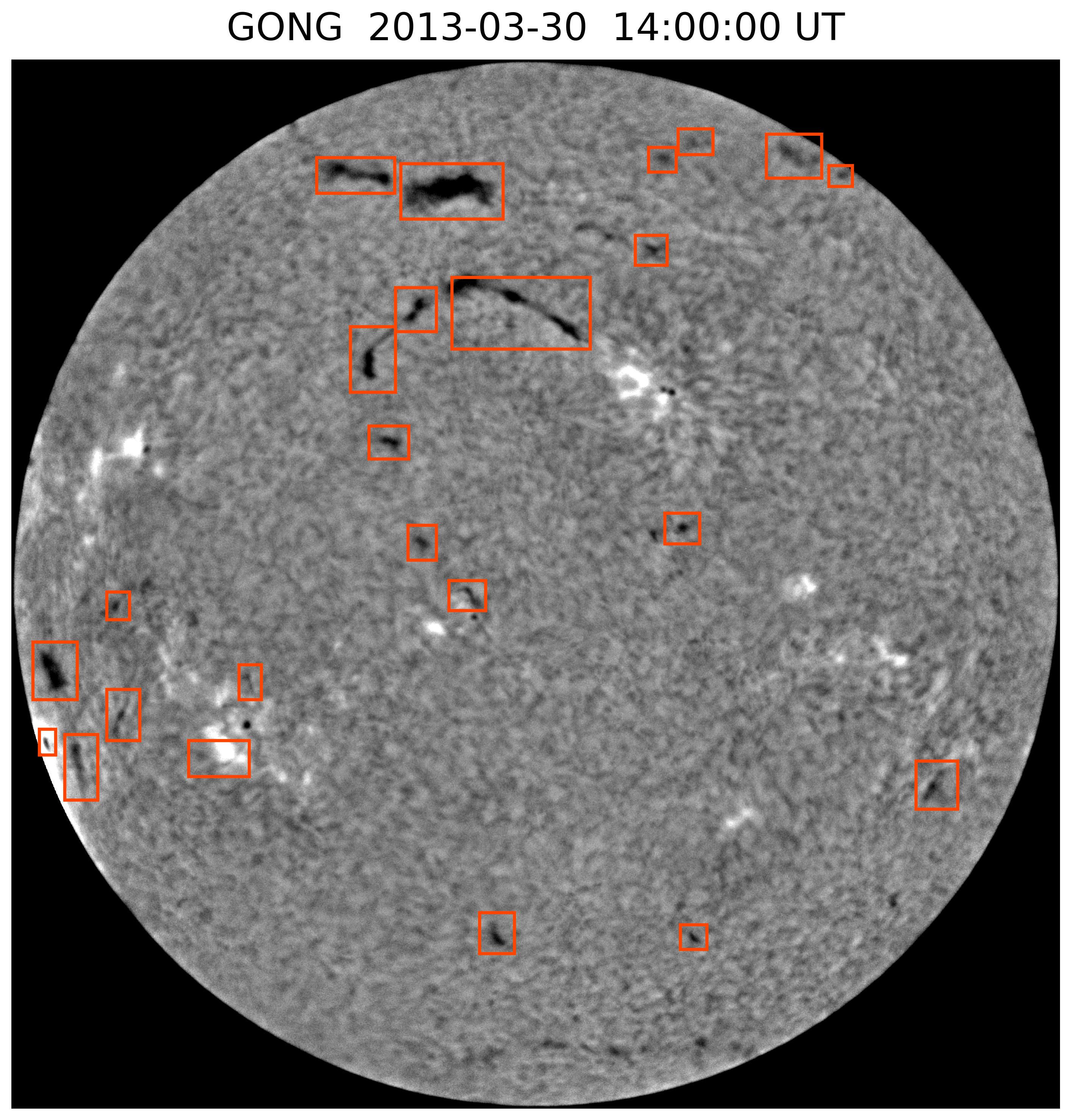}
\caption{Left: Comparison of the ground truth bounding boxes (blue) with predictions of YOLO  (red box) for ChroTel H$\alpha$ filtergrams. Right: YOLO predictions on  GONG H$\alpha$ filtergrams.} 
\label{fig:comp_yolo}
\end{figure*}

Evaluating the accuracy, the highest accuracy is found for the u-net segmentation with a median value (50th percentile) of 94\%. The accuracy is comparing all ground truth bounding boxes with the number of false negative (FN) detection, this is where we have a ground truth bounding box, but no bounding box in the predictions. This means in most cases the u-net segmentation predicts a filament where we have a filament in the ground truth. The precision is low, with only a mean value of about 50\%, which could result from a large number of false positives (FP). False positives are defined as bounding boxes with an IoU value below the threshold of 0.1. In this, the bounding boxes which are too small and which are overlapping partially with other boxes play a role here. Among false positive detections are also predicted filaments, which are not present in the ground truth. The recall values of the u-net segmentation is very high with mean values of 93\%. This shows that there is a very low number of false negative detections again, similar to the accuracy. The difference between accuracy and recall of the results for morphological image processing is the much smaller contours which are often predicted. In the segmentations with morphological image processing often only a small part of the filament is detected. Then, there is a predicted bounding box within the bounding box of the ground truth, but with a very low IoU value, so that for these cases are not accounted for in the calculation of the recall values. This is especially the case when the opacity within the filament is changing. The detection with neural networks accounts for these cases much better than morphological image processing. We note that this evaluation compares the ground-truth bounding-boxes with the pixel-wise segmentation of our model, and therefore only provide limited information about the pixel-wise segmentation and can be influenced by detections of small scale filaments (see small precision and Fig.~\ref{fig:rsize_thres}). 

In the following we examine the influence of small-scale filaments, which are predicted by the segmentation model, but are not labeled in the ground truth. In Figure~\ref{fig:rsize_thres}, we calculate accuracy, precision, and recall for thresholds for the bounding box areas from 2~square pixels to 100~square pixels using the predictions on the test set. In these cases, only bounding boxes with areas above the given threshold  for the ground truth and the predictions by u-net are taken into account for the evaluation. Accuracy and recall decrease with higher thresholds for the bounding boxes. The precision increases mean values of about 75\% for a threshold of 100~square pixels. This shows the large influence of additional predicted small-scale filaments, which are otherwise categorized as false positives.

\subsection{Filament Detection with YOLO} \label{sec:yolo}

All of the shown examples in the following sections from ChroTel and GONG are filtergrams from the test set. In the example in Figure~\ref{fig:comp_yolo} (left) from 2013~March~30 we compare the ground-truth bounding boxes (blue rectangles) and the predicted bounding boxes (red rectangles). Here,  most filaments are detected by YOLO. The scattered polar crown filaments at the southern pole are detected individually as the ground truth is constructed.

On the one hand, there are single filaments with very low opacity, which are labeled in the ground truth, but not recognized by YOLO, but on the other hand there are also filaments, which are missed in the ground truth, but correctly detected by YOLO.  In some cases, YOLO detects filaments, which were missing in the ground-truth. The precision in detecting large-scale filaments seems to be visually higher, but also small scale filaments are well detected. In Figure~\ref{fig:comp_yolo} (left) there are some small-scale filaments at disk center which are not predicted by YOLO, but were labeled for the ground truth. During the labeling of the filaments, filament structures with gaps in between were labeled as individual filaments. This behaviour is reproduced by the YOLO detections, especially visible for polar crown filaments, which are often visible as scattered small-scale filamentary structures. In Figure~\ref{fig:comp_yolo} (left), a filament was labeled individually in the ground truth, but the detection algorithm detected the filament as a single structure.  Furthermore, thin clouds in the data are successfully not identified as filaments (not shown). The results also suggest that the YOLO detection algorithm effectively omitted sunspots from predictions, with bounding boxes aligning closely with the detected filaments in the majority of cases.

The labeling with bounding boxes is conveniently fast, however the separation of filaments is subjective, and quantities, for example, the total area and shape of the filament are of primary importance. For this reason, we use an semi-supervised learning approach and utilize the bounding boxes to obtain pixel-wise segmentation maps. We apply the YOLO algorithm to the ChroTel data of the years 2019 and 2020, which are not labeled, and to the entire GONG data set from 2010 to 2021. The complete ChroTel and GONG data set is used for the training of the u-net segmentation algorithm.  An example of the YOLO predictions on GONG data is shown in Figure~\ref{fig:comp_yolo} (right). The filaments with large opacity are very well detected. Because of the differences in the spatial resolution of GONG and ChroTel data, small-scale filaments and filaments with less opacity are sometimes missed in GONG data. The polar crown filaments in the southern hemisphere are entirely missed in the example in Figure~\ref{fig:comp_yolo} (right), but they are also less well visible as in the ChroTel H$\alpha$ filtergrams. In contrast to the ChroTel data set, where we selected only the best image of each day, the GONG data set contains images with stronger seeing effects. In the shown example, the sunspots are successfully excluded in the detection.

One further difference between the YOLO predictions on ChroTel and GONG data is the prediction of the large central filament in Figure~\ref{fig:comp_yolo}. In ChroTel the filament is labeled and predicted as one large filament, whereas in GONG the filament is detected as three different filaments. This is most likely due to the seeing effects in the GONG image and the different sensitivity of opacity in both instruments, where the filament in GONG appears less opaque and therefore these parts could be mistaken as a separation between different filaments by YOLO.

\subsection{Segmentation of Filaments Using u-net} \label{sec:unet}

\begin{figure*}
\includegraphics[width=0.49\textwidth]{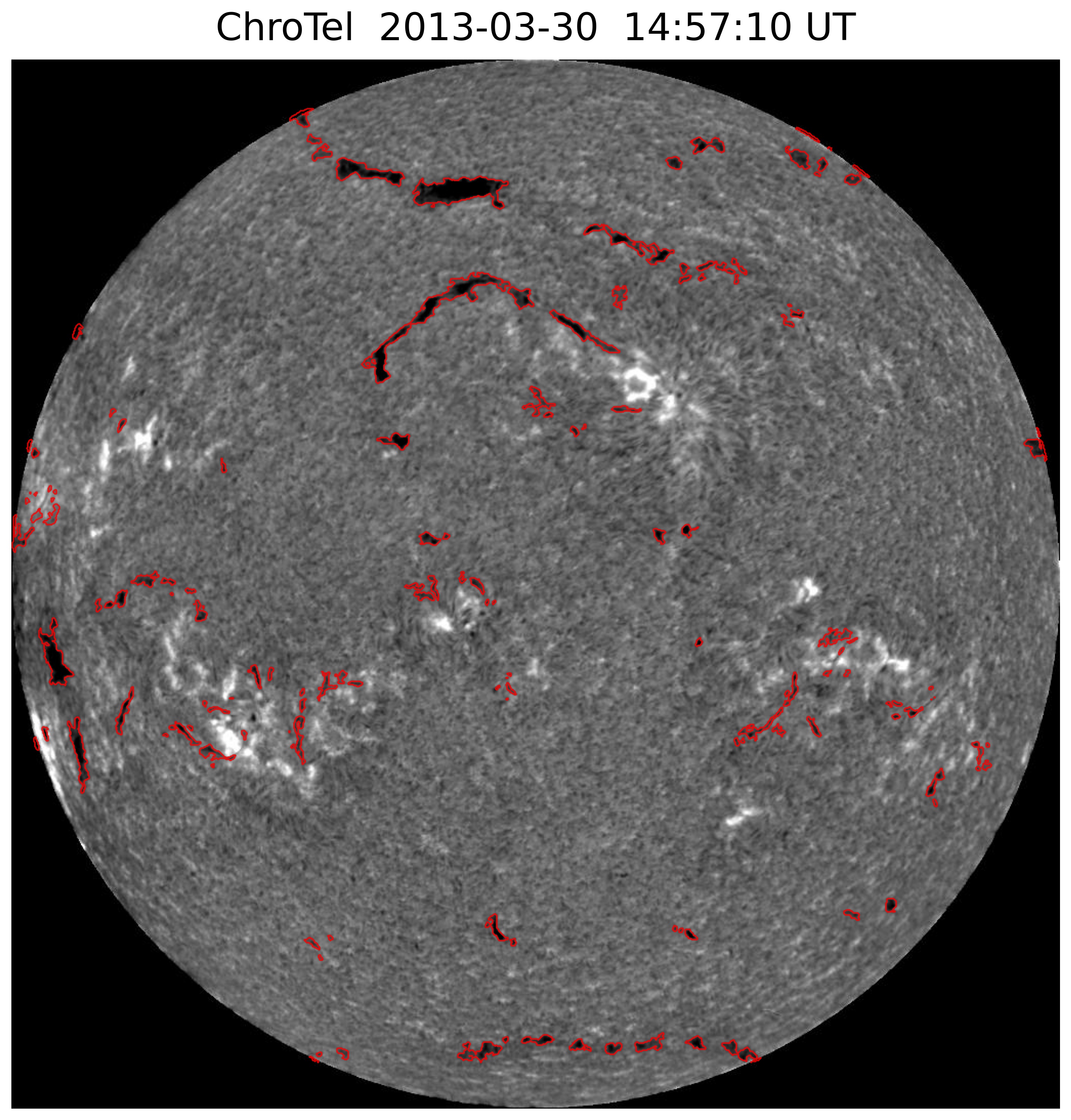}
\includegraphics[width=0.49\textwidth]{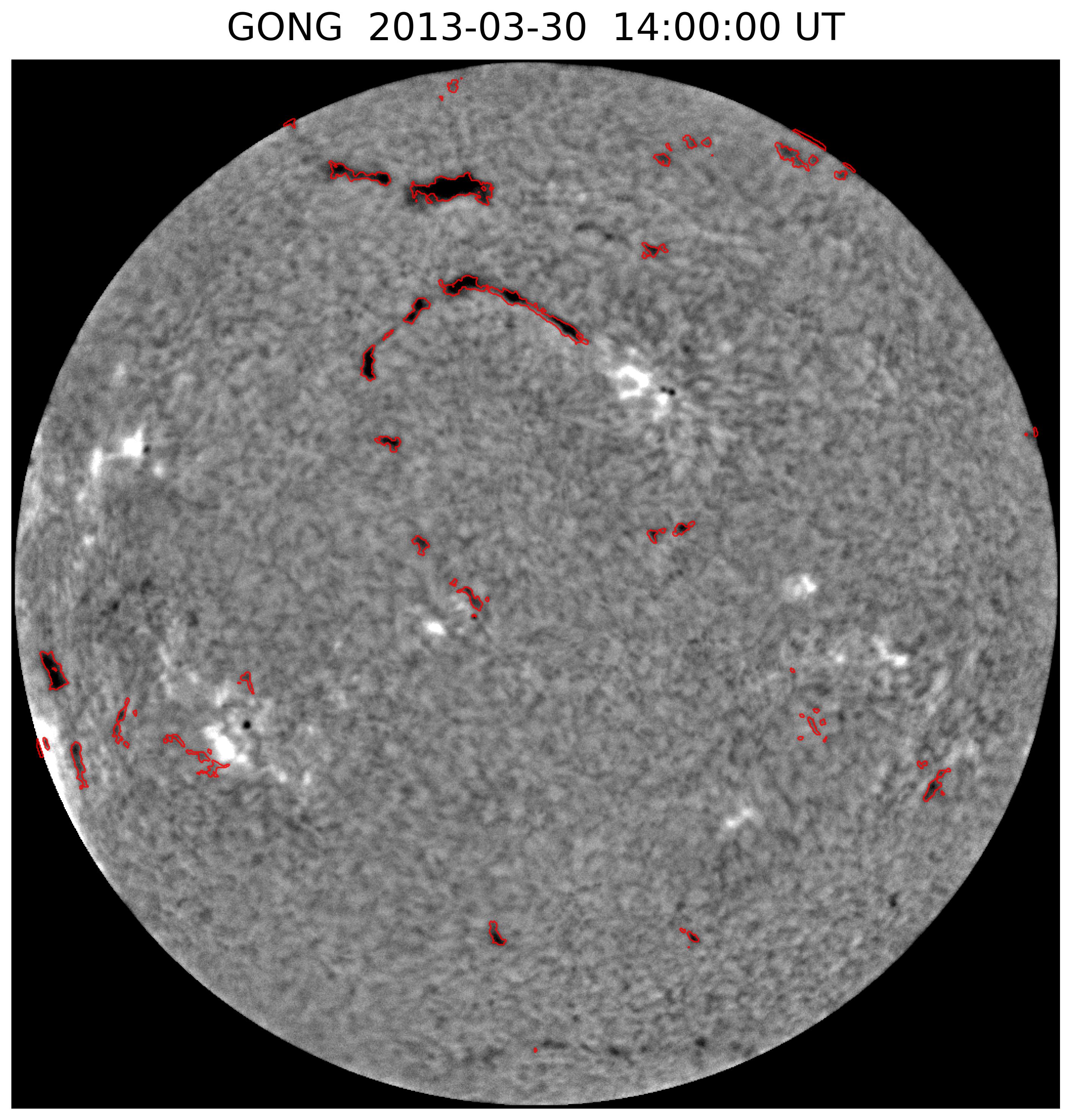}
\caption{Example for H$\alpha$ filtergrams from ChroTel (left) and GONG (right) with the corresponding segmentation maps from u-net as red contours.}
\label{fig:comp_seg}
\end{figure*}

\begin{figure*}
\includegraphics[width=1\textwidth]{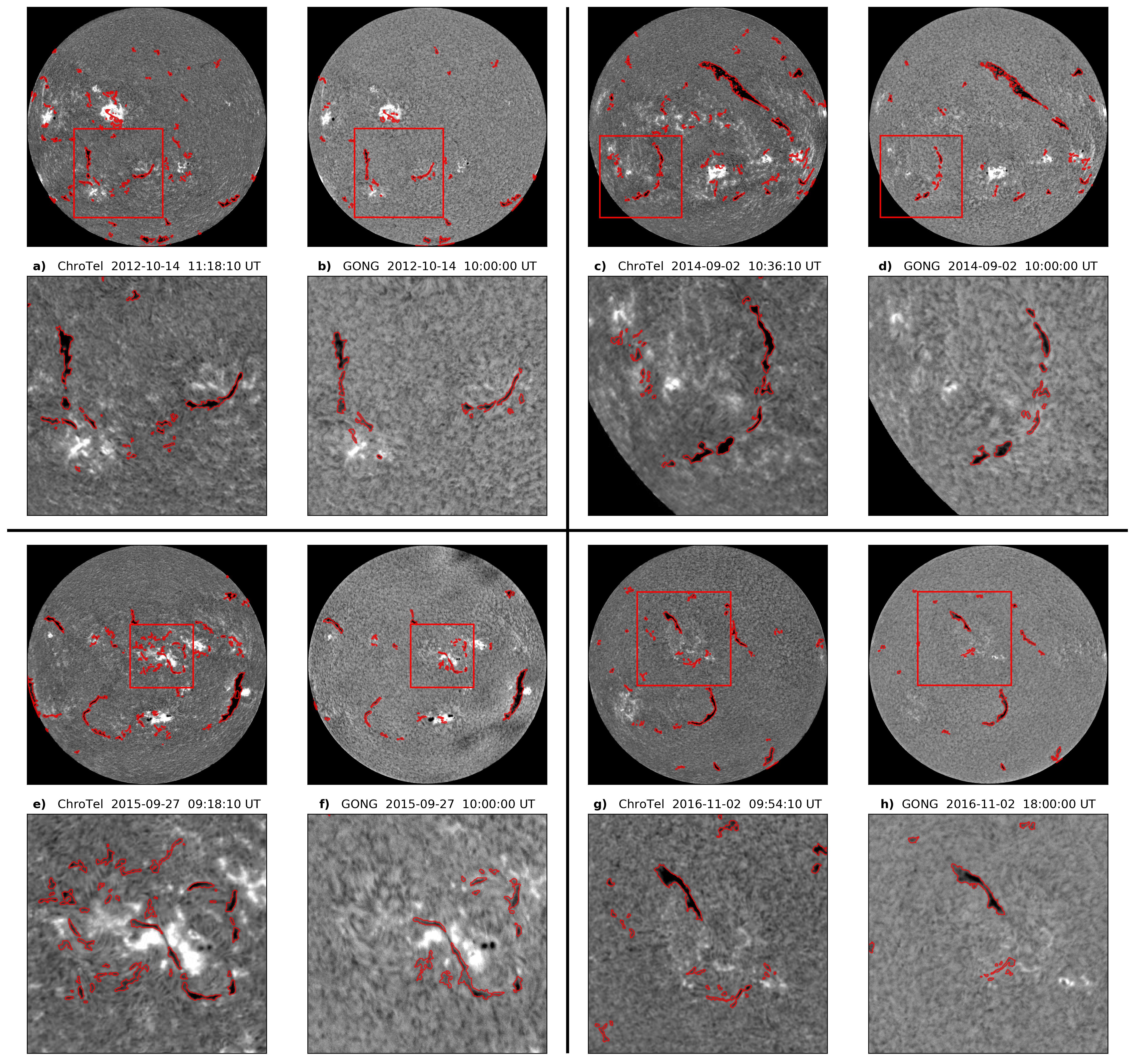}
\caption{Zoom-in in selected segmentation results with u-net for ChroTel (a, c, e, g) and GONG (b, d, f, h) with the contours of the segmented filaments with u-net (red).}
\label{fig:seg_zoom}
\end{figure*}

Example segmentation maps for ChroTel and GONG  are displayed in Figures~\ref{fig:comp_seg}. 
We selected the same examples as for the YOLO object detection in Figure~\ref{fig:comp_yolo}. Most of the filaments are well detected and segmented. All large-scale filaments are detected, but some small-scale filaments are missed. By sighting the examples from the test data set, the segmentation of filaments shows better results than the object detection with YOLO. In Figure~\ref{fig:seg_zoom}, we show further examples of segmented maps with selected  regions-of-interest (ROIs) of segmented active region filaments and quiet-Sun filaments. For active regions, the segmentation on the detail-rich ChroTel filtergrams is very sensitive that sometimes penumbral filaments are detected as active region filaments. For quiet-Sun filaments, the scattered filament parts are well detected and can be nicely used e.g., for tracking the filament in time (see Sect.~\ref{sec:track}).

\begin{figure}
\includegraphics[width=1\columnwidth]{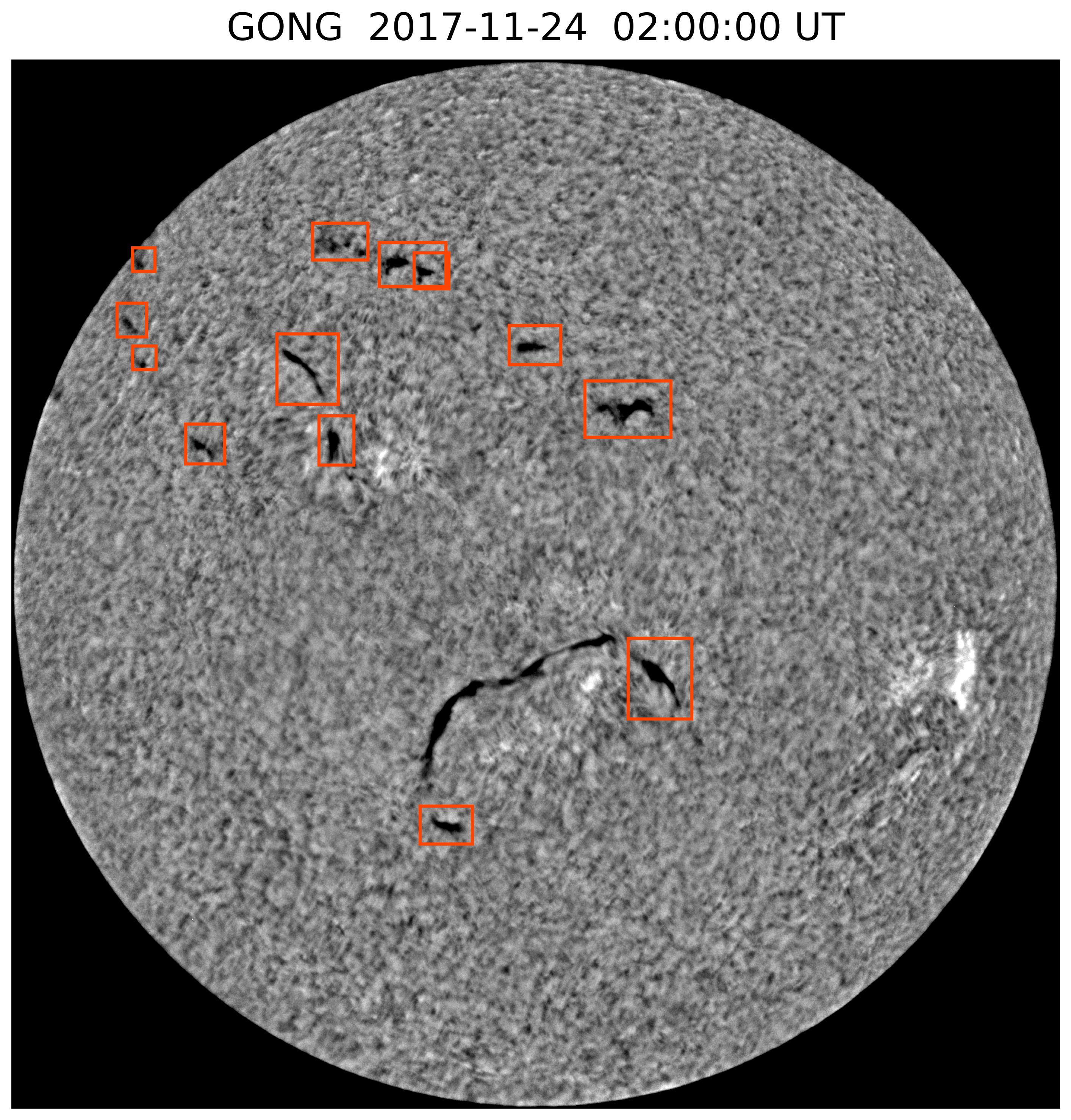}
\includegraphics[width=1\columnwidth]{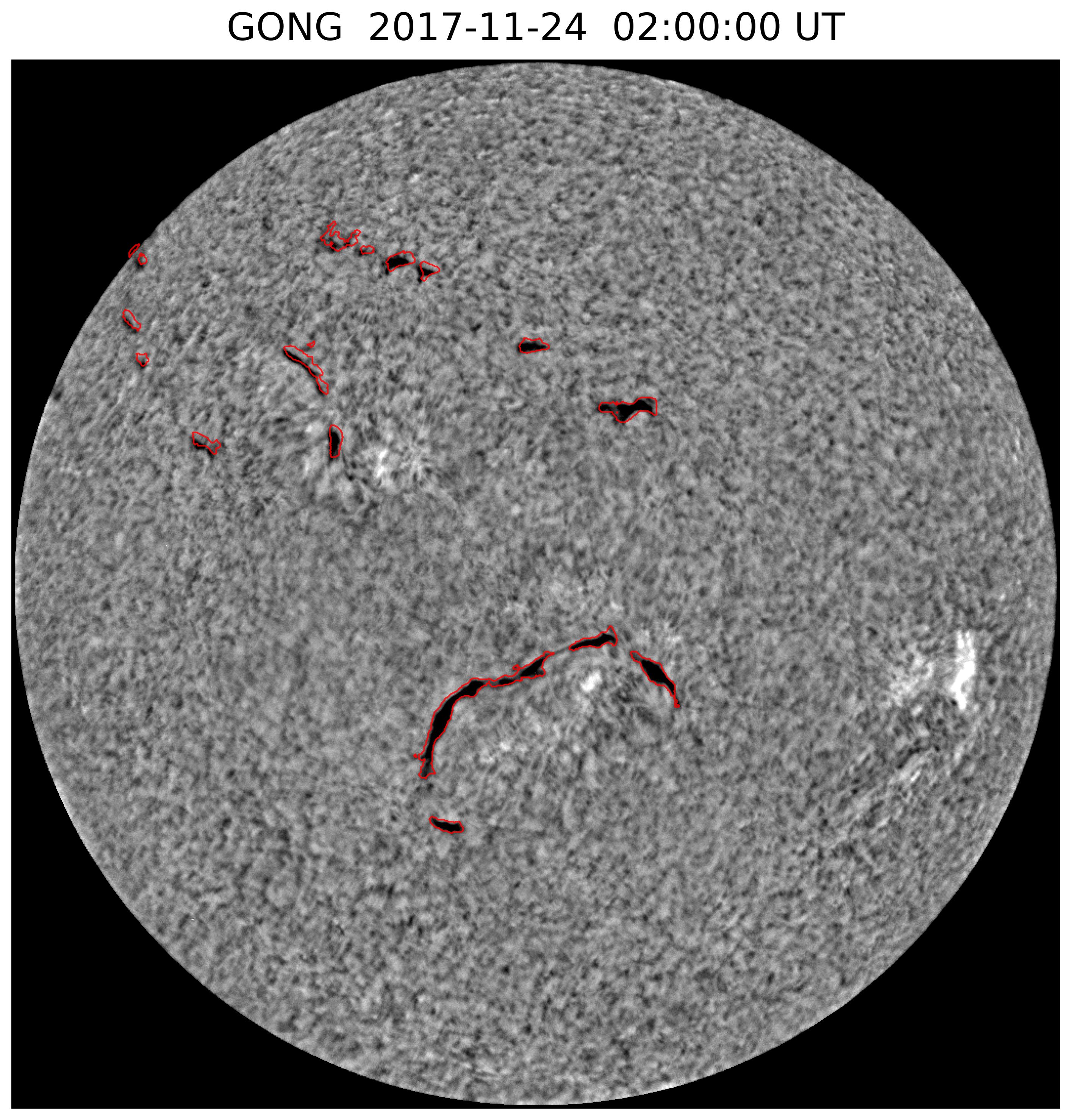}
\caption{Comparison GONG results with YOLO (top) and u-net (bottom).}
\label{fig:comp_gong}
\end{figure}

For the segmentation of GONG, most of the filaments are successfully detected. Figure~\ref{fig:comp_seg} (right) shows a sample segmentation map of GONG. As for the YOLO detection in Figure~\ref{fig:comp_yolo} (right), large-scale filaments are detected, but small-scale filaments are occasionally neglected. There are even some small-scale filaments at the northern pole detected, which are missed in the YOLO detection. Especially for poor seeing conditions, the segmentation of GONG data with u-net leaves out small-scale filaments. In Figure~\ref{fig:seg_zoom}, we display further examples of segmented GONG data with selected ROIs of the images. Also here the shape of the filaments are nicely preserved and represented, especially for quiet-Sun filaments. In GONG filtergrams there is less H$\alpha$ fine-structure visible, nonetheless, sometimes even here some penumbral filaments are mistaken for filaments. For the example on 2014~September~2 (Figure~\ref{fig:seg_zoom}, d), the main quiet-Sun filament is entirely detected, but on the left side, there are some small-scale filamentary structures, which are detected in the ChroTel filtergram in Figure~\ref{fig:seg_zoom} (c), but not in the GONG filtergram.  The segmentation algorithm with u-net is able to deal with appearing clouds as for 2015~Sept~27 (Fig.~\ref{fig:seg_zoom}, f). In Figure~\ref{fig:comp_gong} (top), the large quiet-Sun filament is only partly detected by YOLO, but the segmentation with u-net nicely segmented the entire filament (Figure~\ref{fig:comp_gong}, bottom). This shows that the increase of the training data for the u-net increased also the reliability of the segmentation. Since YOLO is only trained on ChroTel data, the YOLO detections are not as stable as the segmentation results with u-net.

\subsection{Comparison with segmentations from morphological image processing}

\begin{figure}
\includegraphics[width=1\columnwidth]{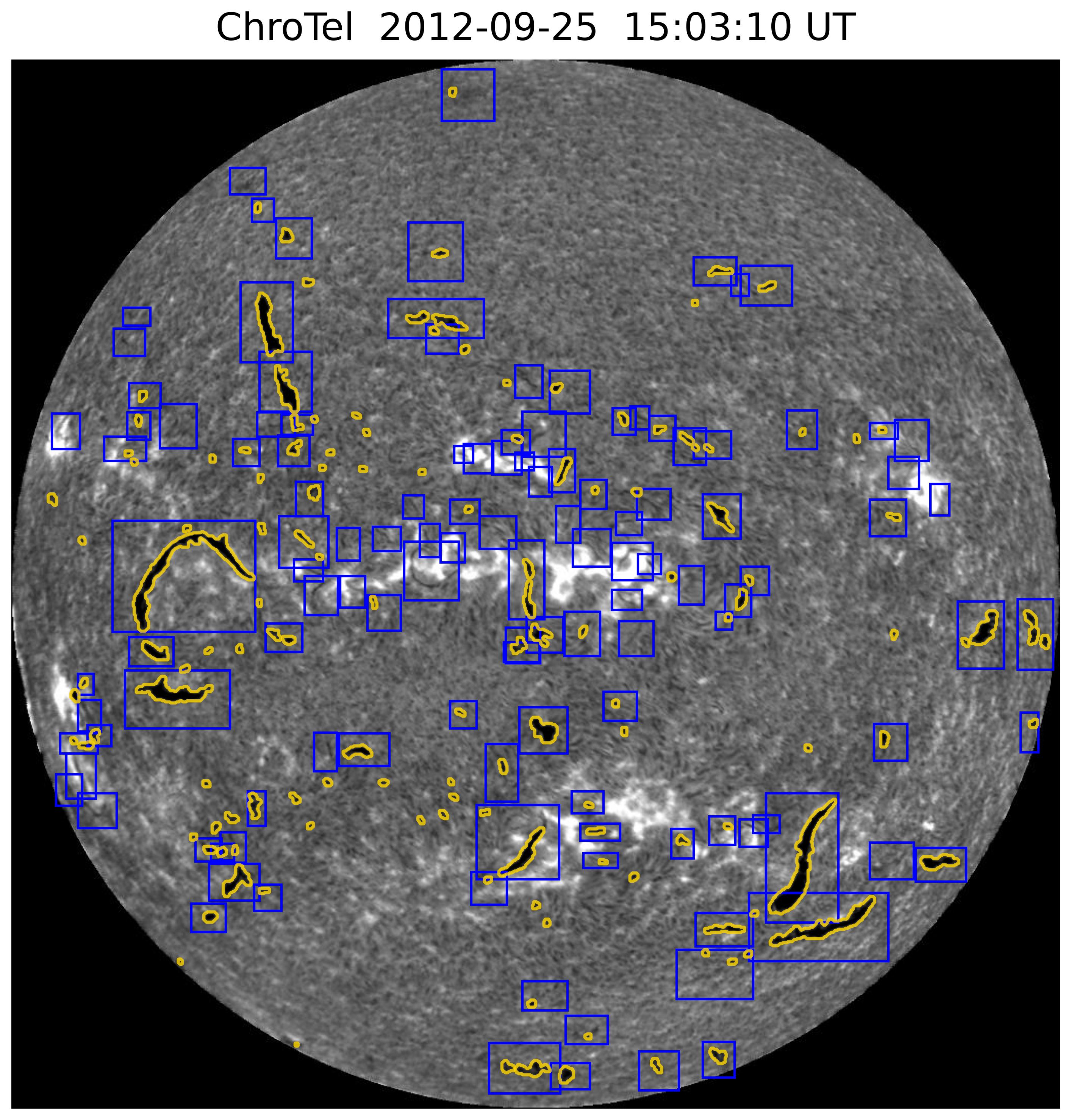}
\includegraphics[width=1\columnwidth]{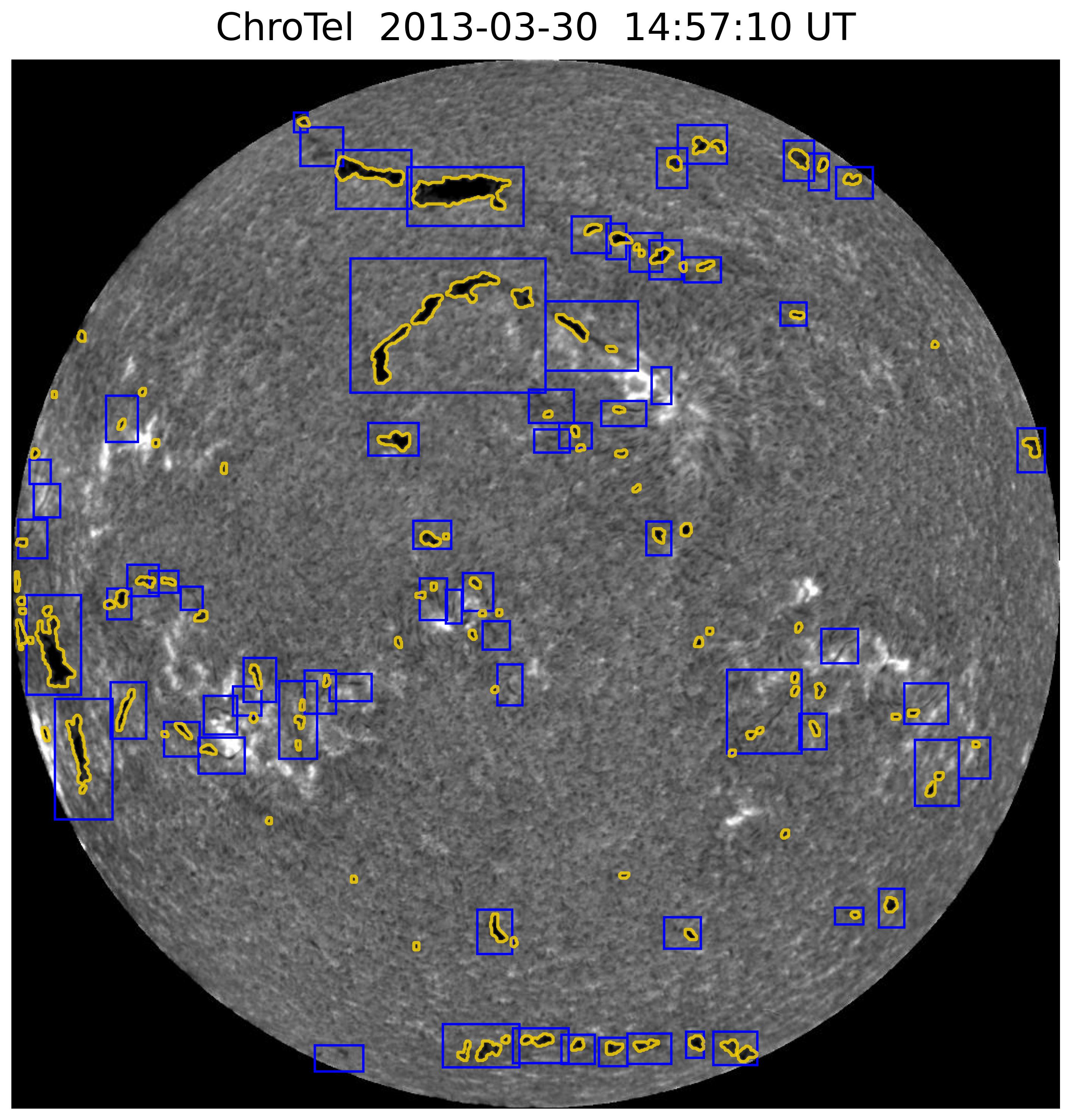}
\caption{Comparison of a segmentation maps created with morphological image processing (yellow contours) with the ground truth (blue boxes) for two different day.}
\label{fig:comp_morph}
\end{figure}

\begin{figure*}
\centering
\includegraphics[width=0.32\textwidth]{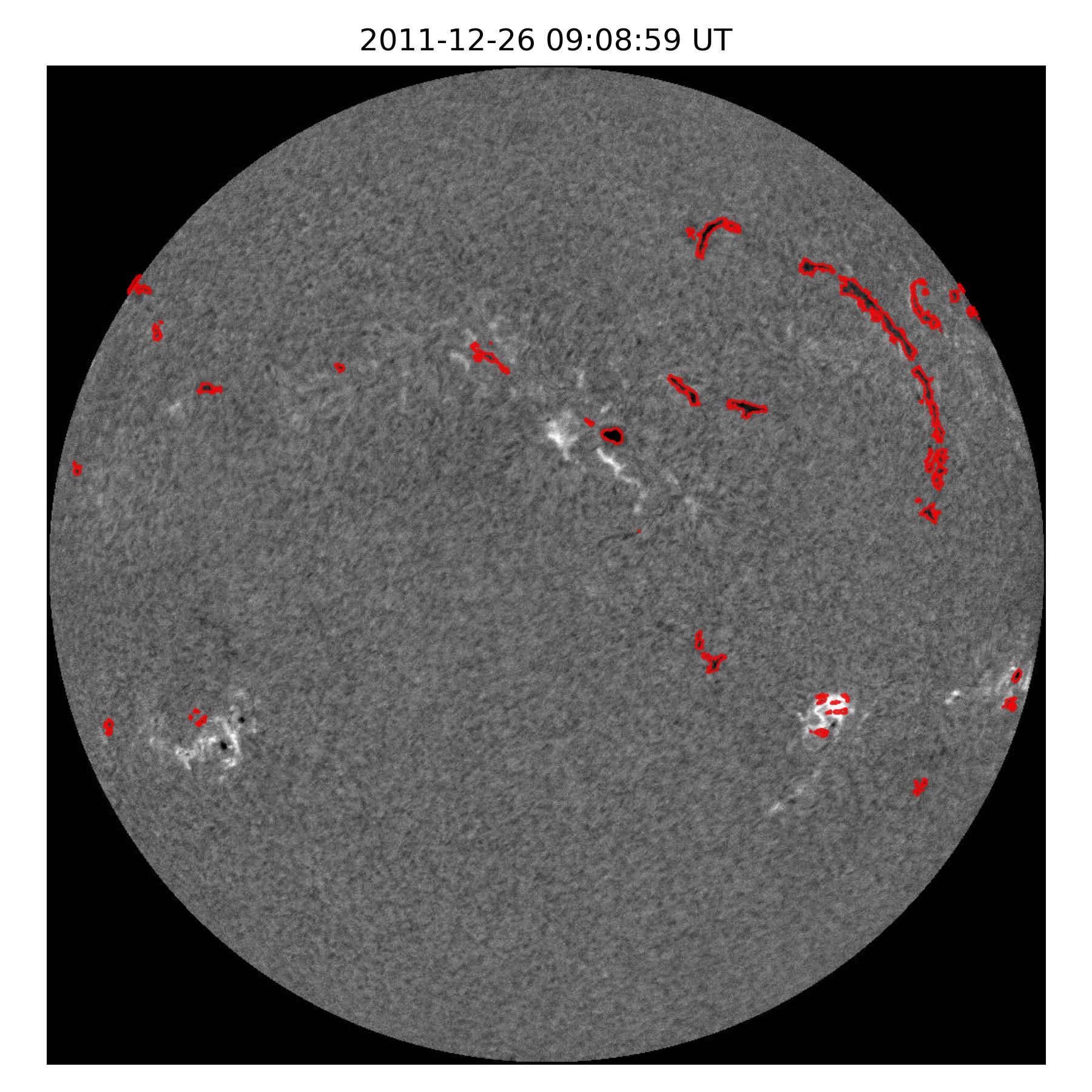}
\includegraphics[width=0.32\textwidth]{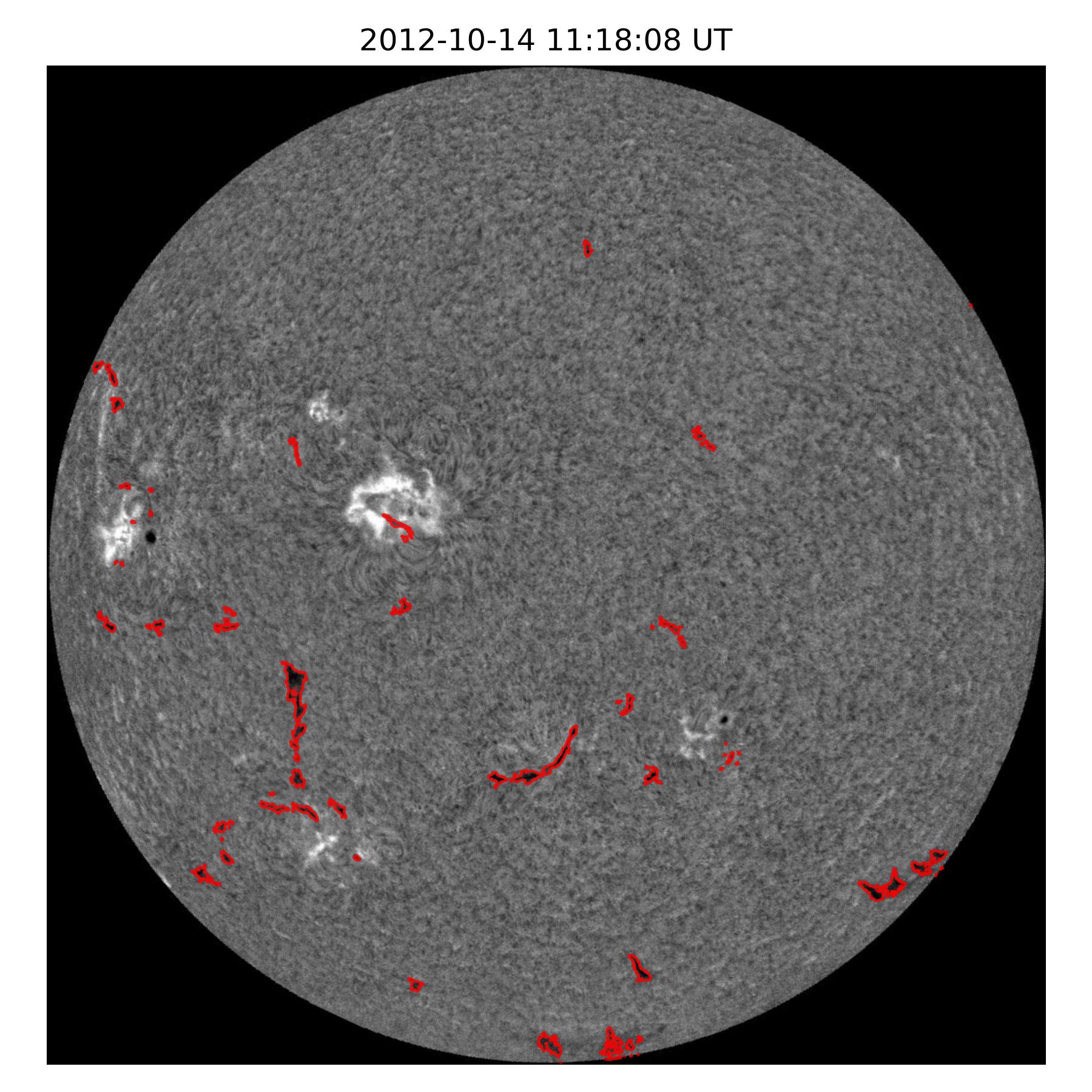}
\includegraphics[width=0.32\textwidth]{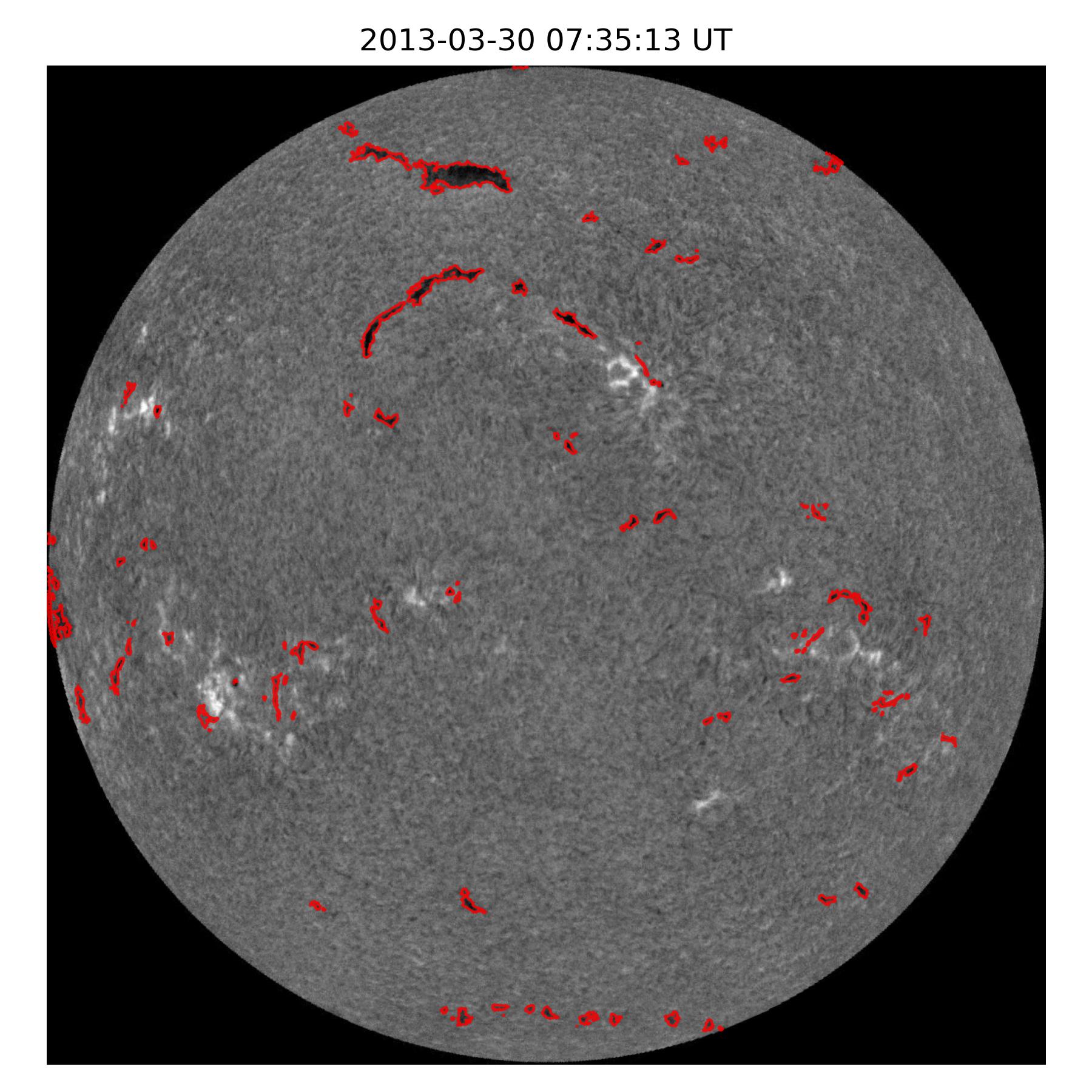}
\includegraphics[width=0.32\textwidth]{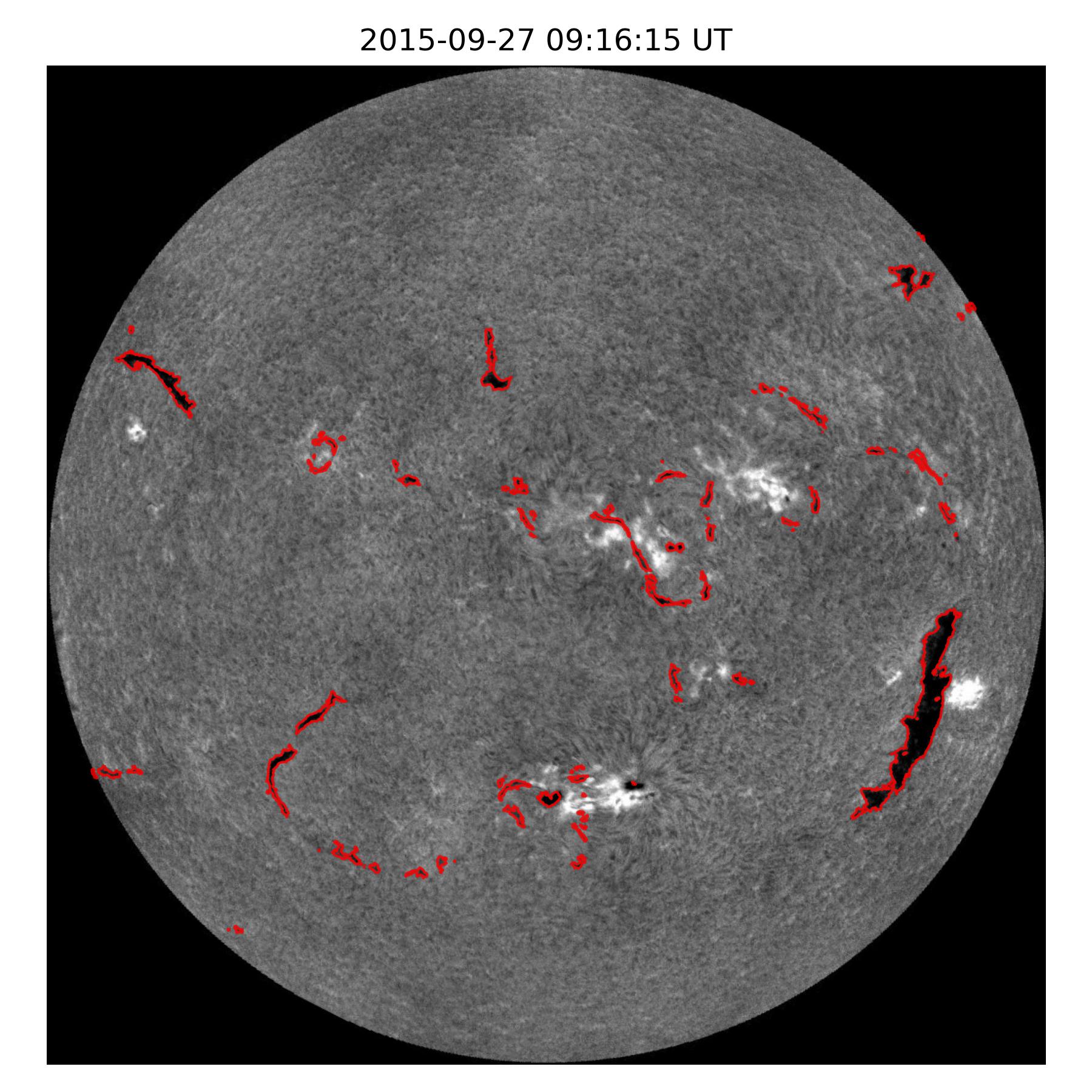}
\includegraphics[width=0.32\textwidth]{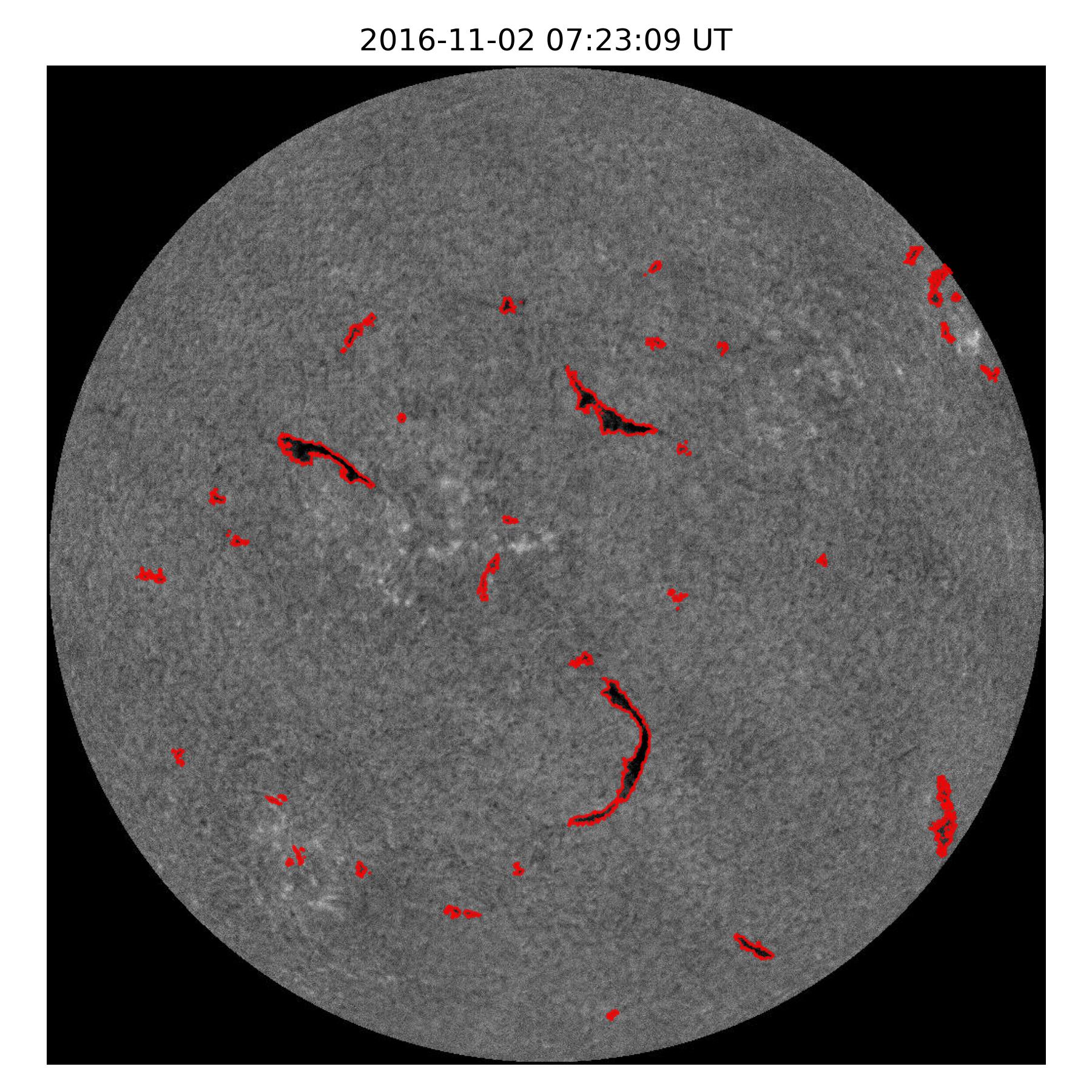}
\includegraphics[width=0.32\textwidth]{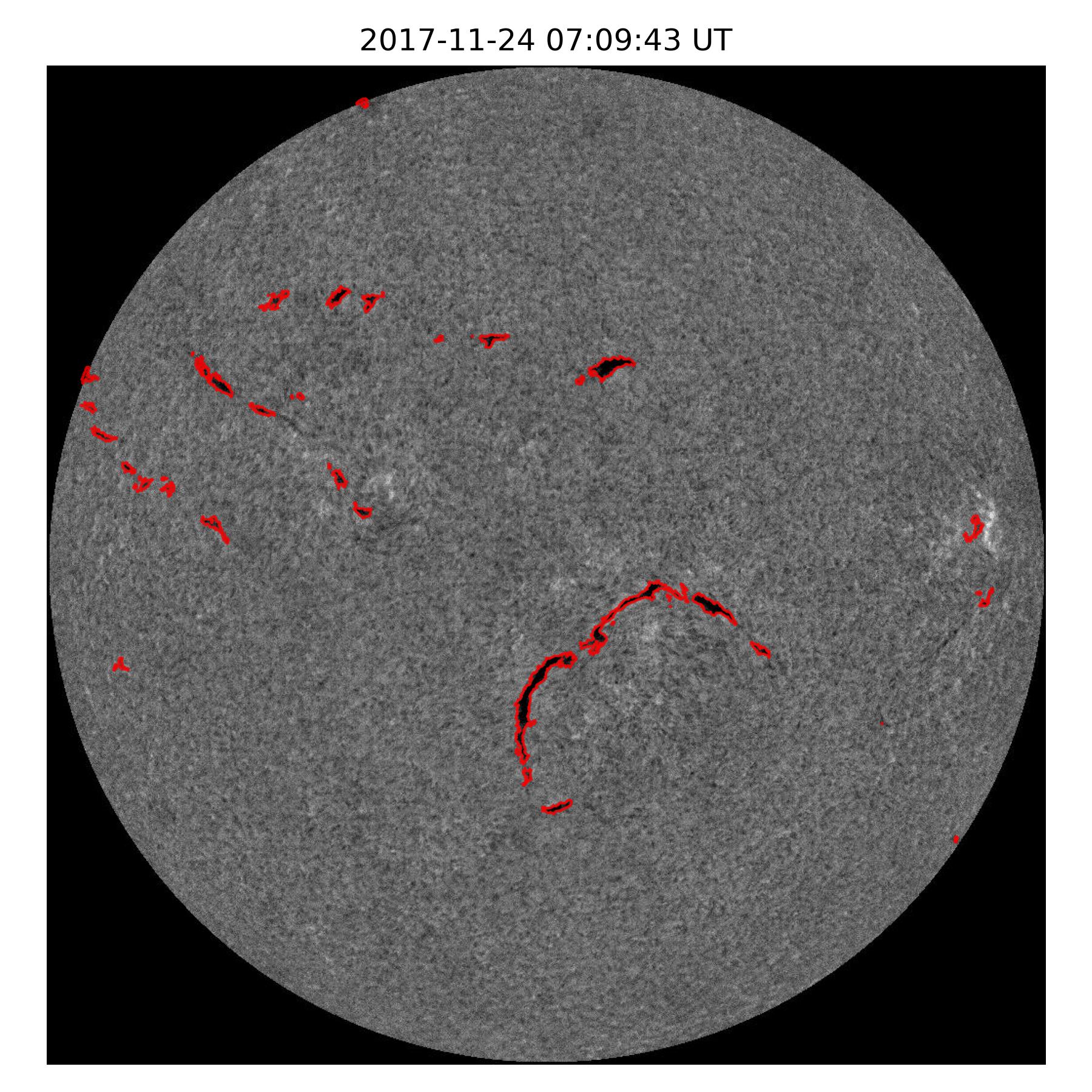}
\caption{Segmentation results from u-net for KSO images with the contours of the segmented filaments (red) for six examples for different times in the dataset with different solar activity.}
\label{fig:kso_seg}
\end{figure*}

In Figure~\ref{fig:comp_morph}, we display a comparison of the segmentation maps created with morphological image processing (yellow contours) and compared to the ground truth (blue boxes).  Larger filaments with high opacity are very well detected by morphological image processing, but if the filament has different opacities, the classical algorithm misses parts of the filament or detects the filament as two structures, where in the ground truth just one filament is labeled. Furthermore, the thresholding method struggles with filaments in between active regions, which have a higher intensity level. In these regions, more often filaments are missed. Often small-scale filamentary structures are detected as filaments, where no filaments are located. There are a lot of small contours scattered all over the solar disk. Furthermore, morphological image processing has more problems with artifacts at the solar limb.

\section{Application on KSO images} \label{sec:kso}

From the training with data from different instruments, we expect that our neural network is robust to instrumental variability, and can be applied to similar H$\alpha$ observations. We perform a qualitative evaluation with unseen H$\alpha$ data from KSO. The KSO data is processed analogously to our previous data preparation. The results of the segmentation are displayed in Fig.~\ref{fig:kso_seg} for a sample of six filtergrams. Large scale filaments are very well detected as well as most small-scale filaments. Similar to GONG observations, there are small-scale filaments, which are not detected or other dark features, which are falsely detected by the algorithm. For example, on 2015~September~27 in Fig.~\ref{fig:kso_seg} there are clouds in the image, which are successfully omitted by the algorithm. Nonetheless, there are still cases, where sunspots are detected.

Comparing the KSO images in detail with ChroTel and GONG images, we also see here slight changes in the detections. As for GONG, the KSO examples presented here suffer from seeing effects, which influence the detection of small-scale filaments or effecting the detection of continuous filaments as one structure. This is the case for the example on 2013~March~30 (Fig.~\ref{fig:kso_seg}, upper right corner), where only small parts of the filament are detected. As described in Sect.~\ref{sec:yolo} for GONG images, also here the opacity is lower than compared to the ChroTel images (see also Figs.~\ref{fig:comp_yolo} and \ref{fig:comp_seg}), which could be also an effect of the different H$\alpha$ filter properties. In the same example, the southern polar crown filaments are well segmented in the ChroTel (Figs.~\ref{fig:comp_seg}, left) and KSO data (Fig.~\ref{fig:kso_seg}), but not in the GONG example (Figs.~\ref{fig:comp_seg}, right), which could be due to strong seeing effects in the GONG example.

\section{Sample applications of the filament segmentation}

\subsection{Tracking of a filament during the day} \label{sec:track}

\begin{figure*}
\includegraphics[width=1\textwidth]{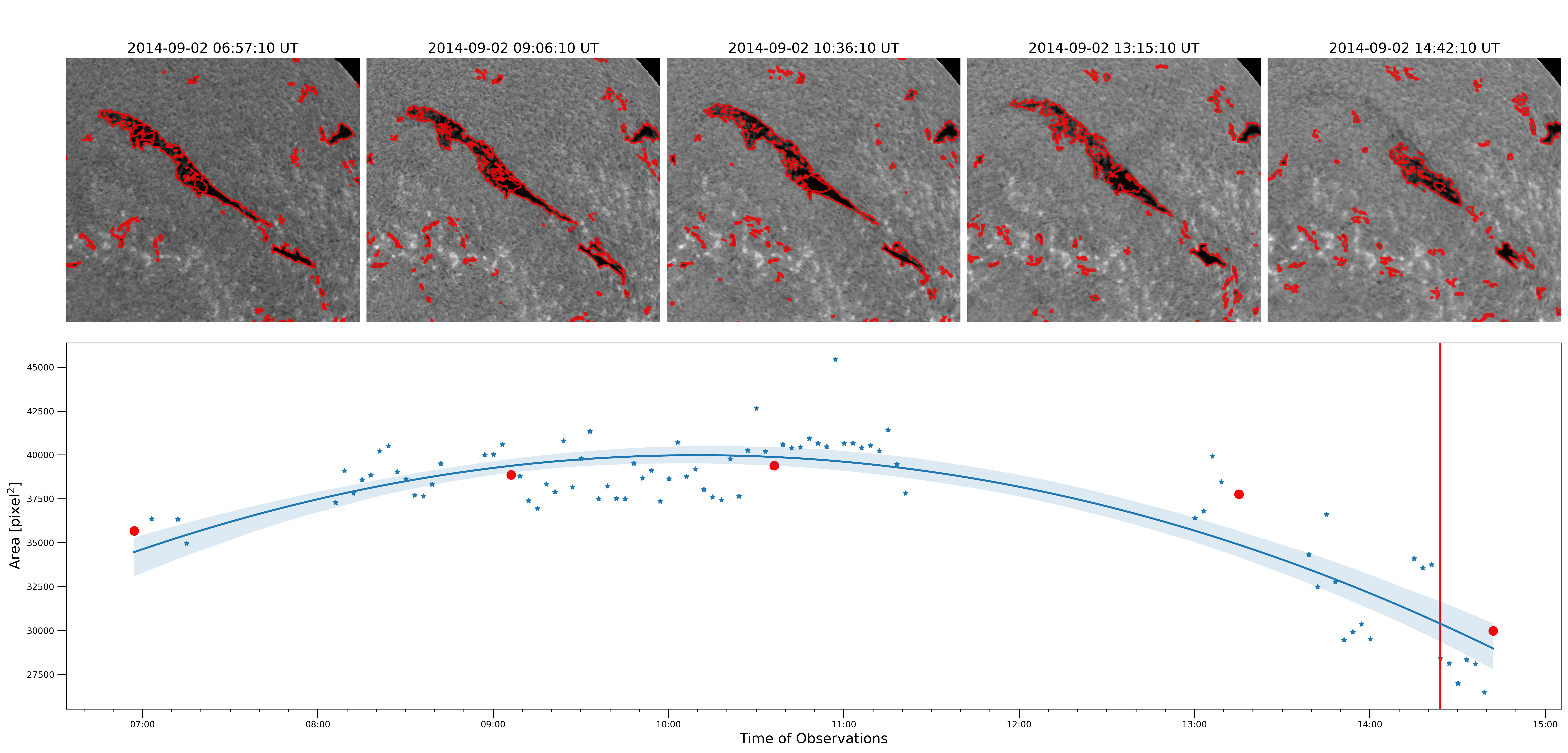}
\caption{Top panels: Evolution of a giant filament on 2014-09-02 from a stable phase until it's eruption with ChroTel H$\alpha$ filtergrams, tracked with the segmentation algorithm (red contours). Bottom panel: Area in pixel$^2$ of the erupting filament for the observing period with ChroTel. The red circles indicate the observing times of the images in the top panels. The red vertical line indicates the start of the eruption. The blue line indicates a quadratic regression through the data points.}
\label{fig:evolv_fil}
\end{figure*}

In order to test the segmentation algorithm, we apply it on a time-series of an erupting filament on 2014~September~2, which is an example from the test set. The time series consists of 89  H$\alpha$ observations from ChroTel taken over the day with a cadence of 3\,min, although there are some gaps in the data due to bad weather conditions. In Figure~\ref{fig:evolv_fil} (top panels), we see the segmentation for a sample of five images. The main filament is well detected, whereby the filament body contains sometimes several contours. The lift-off and eruption of the filament is nicely represented in the segmentation, as well. Nonetheless, in the surroundings, small-scale dark structures are detected, which are not filaments and which are not constantly detected by the algorithm. This sample data set can be used to track a filament over time and study its evolution. In Figure~\ref{fig:evolv_fil} (bottom panel) we display the evolution of the area in pixels of the filament for the same FOV as depicted in Figure~\ref{fig:evolv_fil} (top panels). The area decreases shortly before the eruption starts with the lift-off of the upper part of the filament. Afterwards the area jumps to a much lower value, where it remains until the end of the observations. 

\subsection{Rush-to-the-Pole} \label{sec:rushpole}

\begin{figure*}
\includegraphics[width=1\textwidth]{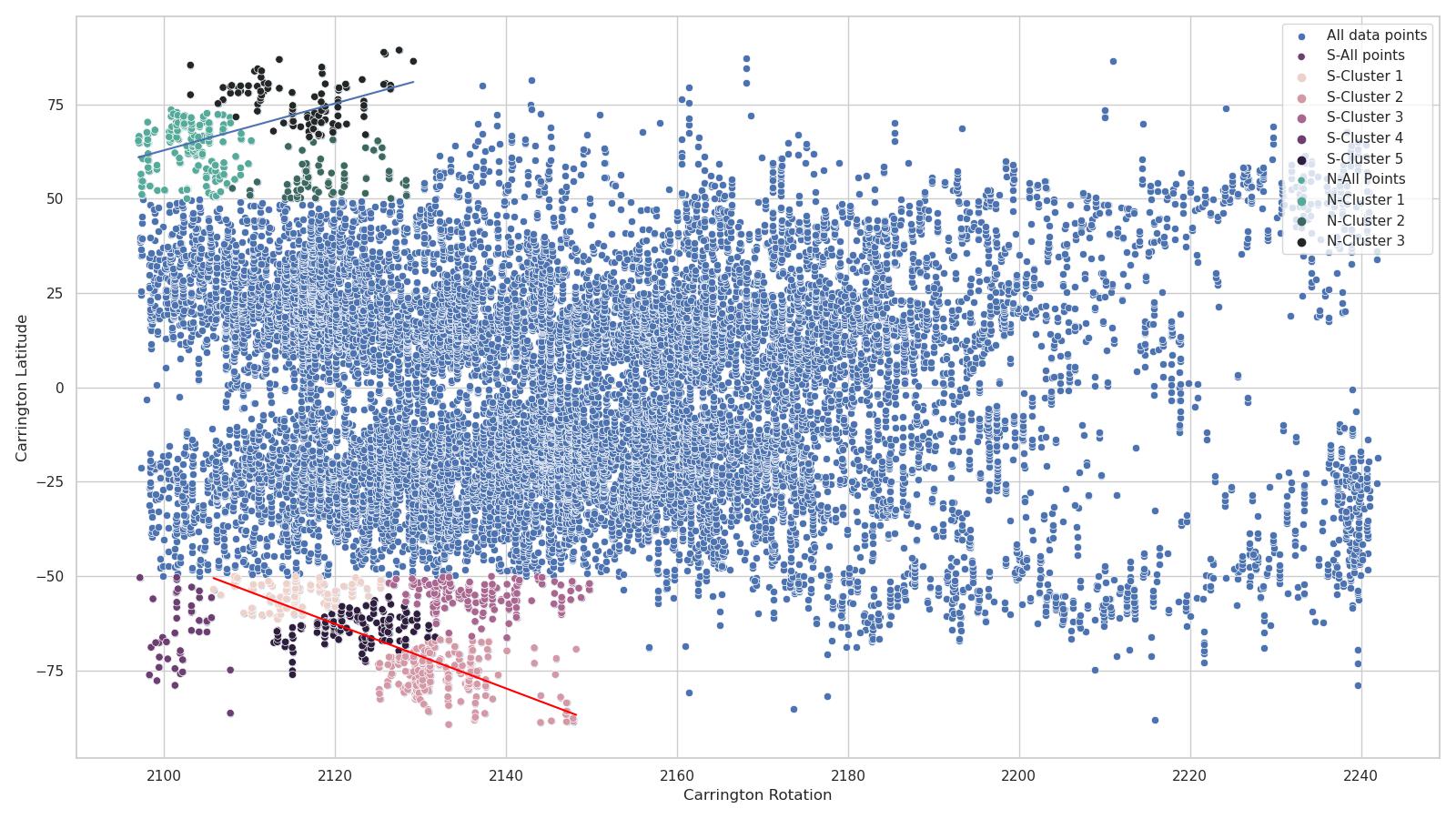}
\caption{Scatter plot of detected filaments in the GONG data set (2010-2021) for Solar Cycle~24 (blue dots). We cluster the polar regions before the magnetic field reversal with k-Means and determine the rush-to-the-pole for northern and southern hemisphere (blue and red line, respectively).}
\label{fig:rushpole}
\end{figure*}

Segmentation maps of filaments can be used to create synoptic Carrington maps \citep{Carrington1858, Diercke2019b, Chatzistergos2023} in order to evaluate the location of solar filaments throughout the solar cycle. This is done in the study of \citet{Diercke2019b} with ChroTel data using morphological image processing to create the segmentation maps. Here, we present an example of a statistical study to determine the rush-to-the-pole from automatic determined segmentation maps of GONG  between 2010 and 2021. We  create synoptic Carrington maps from the segmentation maps indicating the filament locations for Solar Cycle~24. Figure~\ref{fig:rushpole} displays the location of filaments on the synoptic Carrington map. In this scatter plot, we can determine the cyclic behavior of solar filaments. Around the solar cycle maximum, the number of filaments increases at low latitudes and towards the minimum (right half of the plot) the number of filaments decreases. In the beginning of the Solar Cycle (left half of the plot), we can identify the rush-to-the-pole of polar crown filaments, which appear around the minimum at mid-latitudes around $\pm50\arcdeg$ and appear closer to the poles towards the maximum. With the magnetic field reversal, they disappear from the solar surface.

In order to determine the migration rate of polar crown filaments, we use a similar method as described in \citet{Diercke2019b}, where the data points in each polar region before the magnetic field reversal are clustered and the migration rate is determined by a linear regression through all data points of the selected clusters. In the present analysis, we use the k-Means implementation from the \texttt{Scikit-learn} library for \texttt{Python} for clustering the filament data points. The different clusters are color indicated in Figure~\ref{fig:rushpole}. We select for the northern hemisphere the Clusters 1 and 3 and use linear regression through all data points of the clusters to determine the propagation rate. For the southern hemisphere we use five clusters and we select  Clusters 1, 2, and 5 for the linear regression.  The results for the migration rate $m$ are: $m_\mathrm{North} = 0.61\arcdeg$ per rotation and $m_\mathrm{South} = 0.84\arcdeg$ per rotation, whereby a Carrington rotation is set as 27~days. In \citet{Diercke2019b} the migration rate for the southern hemisphere using ChroTel data is determined with $m = 0.79\arcdeg \pm 0.11\arcdeg$ per rotation.

%
\section{Discussions}  \label{sec:disc}
%

By sighting the results from the qualitative evaluation in Sect.~\ref{sec:eval}, we conclude that the object detection algorithm YOLOv5 can obtain filament detections within seconds with a mean accuracy of 85\%. Filaments are detected at all scales and independent of their location and opacity, including small active region filaments located in bright plage regions or close to the solar poles. Even filaments, very close to the limb and erupting filaments are identified by the neural network (for example see Fig. \ref{fig:seg_zoom}). The main driver for this study was to differentiate between filaments and sunspots, which cannot be trivially solved with morphological image processing as seen in \citet{Diercke2019b}, where we excluded small-scale structures to avoid sunspots in the data, but this approach excluded small-scale filaments as well. There is a large number of polar crown filaments of small-scale which are missed out. Often a larger polar crown filament is visible as many small-scale scattered structures. In order to infer basic physical properties of filaments, e.g, location, length, width, connectivity to other layers; and to perform effective statistical studies with these properties, in particular of polar crown filaments and their migration, an effective localization of filaments is of utmost importance. The segmentation with u-net provides exactly this task. We are able to segment the filament completely despite changes in seeing effects. By plotting the heliographic latitude in form of a butterfly diagram over the evolution of Solar Cycle~24, we demonstrate an easy application for the segmentation and we could determine the rush-to-the-pole of polar crown filaments in both hemispheres (Sect.~\ref{sec:rushpole}).

The YOLOv5 is a well established detection algorithm, which is easily installed, and trained. The detection process is possible in real-time, once the network is trained. This gives a very fast possibility to create a labeled data set with a very high precision of filament detections. The model training was only performed with observations from ChroTel. The evaluation of H$\alpha$ filtergrams from GONG and KSO, shows that our method can be directly applied to any new H$\alpha$ data set without additional labeling efforts or model training. In order to further increase the  performance, a training with different data sources would be beneficial.

The segmentation allows us to extract the filaments directly from the images by using the predicted masks generated by the network. In the study of \citet{Zhu2019}, the trained u-net efficiently segmented filaments, but sunspots were not excluded. The major problem was the use of a semi-manual ground-truth, which contained already sunspots. Other methods were developed to separate filaments and  sunspots using  SVMs \citep{Qu2005}, but this needs an additional training and computation. \citet{Ahmadzadeh2019} used a different approach for the training data with ground-truth masks created from the HEK database, which have an accuracy of 72\% compared to hand-labeled data \citep{Ahmadzadeh2019}. In our approach, we used manually labeled data and created the binary input masks from the labeled data. Sunspots are in most cases excluded, while  small-scale filaments and filaments close to the limb are included in the annotations. This results in an optimal precondition for an effective training of the network for the segmentation of filaments. Our accuracy of the segmentation is very high with a median value of 94\%. The results justify the initial effort in manually labeling.

However, in the detail-rich ChroTel data parts of the H$\alpha$ fine-structure and penumbral filaments are sometimes detected as actual filaments. Also sunspots are occasionally detected. In some cases when a sunspot is detected, it is only partly segmented, in particular, if other filamentary structures of an active region filament are in its proximity. Small-scale filaments are often missed in other methods, because of the lack of resolution in the input data \citep{Zhu2019, Liu2021, Guo2022}. With the performed pre-processing, more small-scale filaments in ChroTel and GONG data can be recognized in the images by the presented segmentation method. Moreover, the segmentation of GONG data is stable for different seeing condition, even if clouds cover parts of the solar disk, and is less sensitive to instrumental differences, as can be seen from the application to KSO data. 

From Fig. 6 it can be seen that with a threshold of about 20 square pixels the precision largely improves to $>60$\%, while recall and accuracy are marginally decreasing ($\approx 90$\%). Therefore, for studies that do not require small-scale filaments, a threshold of 20\,square pixels is appropriate to reduce potential false positives. Nonetheless, we provide all information on detected filaments or filamentary structures, while leaving it to the scientific application of the users to filter out small-scale detections as needed.

The ground truth labels are generated by three different people who slightly vary in the way they labeled the data. Also after agreeing on a labeling strategy, there are cases, in which more small-scale filaments are labeled; the boxes are defined larger than in other cases; or parts of a larger filament are labeled as one single filament, which complicates the comparison of the ground truth bounding boxes with the segmentation results.

%
\section{Conclusions and Outlook}\label{sec:conc}
%

The presented semi-supervised learning approach using YOLO and u-net for a complete segmentation of H$\alpha$ filtergrams with respect to filaments achieves already a good performance. The method can be further improved by: 1) Extending the training data of YOLO and of u-net by including the KSO data set, which is a very long-lasting data set and even contains data of previous cycles, which enriches the variety of the data samples and makes the training more stable. 2) Additional wavelength channels can be used to improve the performance of the segmentation. KSO and ChroTel have observed \mbox{Ca\,\textsc{ii}}-K filtergrams very close in time. It can be used for a better differentiation between filaments and sunspots, because in the \mbox{Ca\,\textsc{ii}}~K filtergrams sunspots are better visible and typically do not display filaments. Some remnant filamentary structures are sometimes hardly visible in the ChroTel filtergrams of the \mbox{Ca\,\textsc{ii}}~K line \citep[Fig.~4 in ][]{Kuckein2016, Diercke2021}.  The  \mbox{He\,\textsc{i}}~$\lambda$10830\,\AA\ of ChroTel give a further opportunity, because the spectroscopic data allow us to use continuum observations, which only show sunspots, whereas the line-core filtergrams display sunspots and filaments, as well as plage regions as dark structures. 3) The HMI intensity data, with a cadence of 45\,s, can be also used for the exclusion of sunspots. False detections can be reduced by further including line-of-sight magnetograms which contain the polarity inversion line (PIL), over which each filament is formed and strong magnetic field concentrations in the photosphere, for example in sunspots. 4) Detecting solar filaments from a video sequence rather than individual images. This could further improve the temporal consistency and mitigates seeing effects.

The filament detection and segmentation approach is very successfully applied to the ChroTel and GONG database and is applied KSO data. The pipeline includes an automatic image reduction standardized to use it on different data sources. The pipeline including u-net trained for filament segmentation is publicly available on GitHub\footnote{\href{https://github.com/adiercke/DeepFilamentSegmentation}{github.com/adiercke/DeepFilamentSegmentation}}. A large-scale data integration from different telescopes will be possible, e.g., BBSO or the Uccle Solar Equatorial Telescope \citep[USET, ][]{Clette2002}. Further applications could be the automatic detection of eruptive filaments in real-time, which is relevant for the forecast of coronal mass ejections (CME) and space weather events, which could hit Earth. The Sun is observed by many telescopes from all over the world in H$\alpha$. Ground-based solar telescopes are restricted to observations during the day.  Automated and reliable detection is required to analyze the data stream from multi-site observatories.

To foster novel research with solar filaments, we will provide catalogs based on the ChroTel and GONG data set. For ChroTel, we apply the trained segmentation method on the described ChroTel data set, i.e, one image per day. The data will be available in the data archive\footnote{KIS Science Data Centre Archive: \href{https://archive.sdc.leibniz-kis.de/}{archive.sdc.leibniz-kis.de}}  of Science Data Centre (SDC) of the Institute for Solar Physics (KIS) in Freiburg, Germany.  For GONG we apply our method to the full 4-hour cadence data set. The data will be available in the Historical Solar Data archive\footnote{Historical Solar Data archive: \href{https://historicalsolardata.org}{historicalsolardata.org}}. The resulting pixel-wise segmentation has a high temporal cadence and high precision on the full-disk H$\alpha$ filtergrams. The catalog of solar filaments enables new studies of solar filaments. For example, statistical studies of filaments over several decades can be faster obtained. Furthermore, the pixel-wise segmentation allows to obtain parameters such as the size or orientation of the filaments. Ultimately, the catalog paves the way towards space weather forecasting of filament eruptions and the study of triggering of such events. 

Moreover, the data can be used as an input for the HEK database, but utilizing not only BBSO data \citep{Ahmadzadeh2019}, but of several instruments with nearly real-time detection. Furthermore, the segmented filaments can be used as input data for another detection algorithm, e.g., to detect coronal holes \citep{Hofmeister2019, Palacios2020, Illarionov2020, Jarolim2021} in EUV data of SDO. Both filaments and coronal holes are visible as dark objects in the solar corona, i.e. in SDO observations at 193\,\AA. Therefore, often filaments are  detected incorrectly  as  coronal holes. If the network learns to differentiate filaments from coronal holes, the detection of coronal holes can be improved as well \citep[e.g., ][]{Reiss2015}. In addition, the training can be extended by not only giving the location of filaments in the EUV maps, but by training the data with H$\alpha$ full-disk observations.

Beside filaments other structures can be detected with these methods. Other classes can be easily added, for example, sunspots, which are omnipresent in full-disk observations of the photosphere and chromosphere. This would require a labeled data set of sunspots to start  the training, which can be created from morphological maps of dark objects, excluding the already detected filaments. Going from full-disk images to high-resolution ground-based observations enables many new  applications of object detection and segmentation approaches based on deep learning. High-resolution observations of the High-resolution Fast Imager \citep[HiFI, ][]{Kuckein2017IAU, Denker2018} at the  GREGOR solar telescope \citep{Schmidt2012} are collected in a large data archive with observations in G-band (3500~images), blue continuum (2100~images), and \mbox{Ca\,\textsc{ii}}~H (2800~images) on about 110 observing days in 2016. The labeling of the data is in progress and implementing object detection with YOLOv8 is planned in the framework of the SOLARNET project. Starting 2021, the world's largest solar telescope, the Daniel K. Inouye Solar Telescope \citep[DKIST, ][]{Tritschler2016DKIST, Rimmele2020} with a primary mirror of 4\,m-diameter, observes regularly the Sun and in the near future the European Solar Telescope \citep[EST, ][]{QuinteroNoda2022} will start operation. A large amount of data will be collected every day with high frame rates and petabytes of solar data.  Deep learning algorithms for classification, object detection, and segmentation will be of utmost importance for the new era of solar observations.

\begin{acknowledgements}
The National Solar Observatory (NSO) is operated by the Association of Universities for Research in Astronomy, Inc. (AURA), under cooperative agreement with the National Science Foundation (NSF). ChroTel is operated by the Institute for Solar Physics (KIS), Freiburg, Germany, at the Spanish Observatorio del Teide, Tenerife, Canary Islands. The ChroTel filtergraph has been developed by KIS in co-operation with the High Altitude Observatory in Boulder, CO, USA. GONG H$\alpha$ data were acquired by GONG instruments operated by NISP/NSO/AURA/NSF with contribution from NOAA. GONG H$\alpha$ data are available at DOI: \href{https://doi.org/10.25668/AS28-7P13}{10.25668/AS28-7P13}. KSO H$\alpha$ data were provided by the Kanzelh{\"o}he Observatory, University of Graz, Austria. This research has received financial support from the European Union’s Horizon 2020 research and innovation program under grant agreement No. 824135 (SOLARNET). RJ acknowledges support by the SOLARNET program. AD acknowledges the SOLARNET mobility program. AAP acknowledges support by NSF grant N0017.103.OG02.UG. CK received funding from the European Union's Horizon 2020 research and innovation programme under the Marie Sk\l{}odowska-Curie grant agreement No 895955. SJGM is grateful for the support of the European Research Council through the grant ERC-2017-CoG771310-PI2FA, by the Spanish Ministry of Science and Innovation through the grant PID2021-127487NB-I00, and by the support  of the project VEGA~2/0043/24. This research has made use of NASA's Astrophysics Data System. The authors acknowledge the use of the Skoltech Zhores cluster for obtaining the results presented in this paper \citep{zacharov2019zhores}. 
\end{acknowledgements}


\end{document}